\documentclass[11pt]{article}

\usepackage{lipsum} 

\usepackage{bm}
\pdfoutput=1
\usepackage{amsmath,amsfonts,amsthm}
\usepackage{relsize}
\usepackage{esdiff}  
\usepackage{booktabs}  
\usepackage{url}  
\usepackage{cleveref}  
	\crefname{equation}{equation}{equations}
	\crefname{figure}{figure}{figures}	
	\crefname{table}{table}{tables}
\usepackage[skip=1.5pt,font=small]{caption}

\usepackage{caption}
\usepackage{bbm}

\usepackage[usenames,dvipsnames,svgnames,table]{xcolor}

\usepackage{graphicx}
\usepackage{hyperref}  
\usepackage[export]{adjustbox}

\usepackage[sc]{mathpazo} 
\usepackage[T1]{fontenc} 
\usepackage{microtype} 

\usepackage{multicol} 
\usepackage[margin={1cm,1.5cm},columnsep=20pt]{geometry}

\usepackage{booktabs} 
\usepackage{float} 
\usepackage{hyperref} 

\usepackage{lettrine} 
\usepackage{paralist} 

\usepackage{abstract} 
\usepackage{abstract} 

\usepackage{titlesec} 
\renewcommand\thesection{\Roman{section}} 
\renewcommand\thesubsection{\Alph{subsection}} 
\titleformat{\section}[block]{\large\scshape\centering\bfseries}{\thesection.}{1em}{} 

\titleformat{\subsection}[block]{\scshape\centering}{\thesubsection.}{1em}{} 

\usepackage{fancyhdr} 
\DeclareCaptionFormat{myformat}{#1#2#3\hrulefill}
\usepackage{authblk}
\makeatletter
\renewcommand\AB@affilsepx{, \protect\Affilfont}
\makeatother
\title{\vspace{-15mm}\fontsize{16pt}{16pt}\selectfont\textbf{Computational model of avian nervous system nuclei governing learned song}} %
\author[1]{Eve Armstrong\thanks{earmstrong@ucsd.edu}}
\affil[1]{Computational Neuroscience Initiative, University of Pennsylvania, Philadelphia, PA 19104} 
\date{(Dated: \today)}
\renewcommand\Affilfont{\itshape\small}
\begin{document}
\maketitle 
\begin{abstract}
The means by which neuronal activity yields robust behavior is a ubiquitous question in neuroscience.  In the songbird, the timing of a highly stereotyped song motif is attributed to the cortical nucleus HVC, and to feedback to HVC from downstream nuclei in the song motor pathway.  Control of the acoustic structure appears to be shared by various structures, whose functional connectivity is largely unknown.  Currently there exists no model for functional synaptic architecture that links HVC to song output in a manner consistent with experiments.  Here we build on a previous model of HVC in which a distinct functional architecture may act as a pattern generator to drive downstream regions.  Using a specific functional connectivity of the song motor pathway, we show how this HVC mechanism can generate simple representations of the driving forces for song.  The model reproduces observed correlations between neuronal and respiratory activity and acoustic features of song.  It makes testable predictions regarding the electrophysiology of distinct populations in the robust nucleus of the arcopallium (RA), the connectivity within HVC and RA and between them, and the activity patterns of vocal-respiratory neurons in the brainstem.
\end{abstract}

\begin{multicols}{2}

\section{INTRODUCTION}
The song motor pathway is an excellent testbed for probing the relationship between neuronal activity and a highly stereotyped and quantifiable animal behavior.  Specifically, a remarkably sparse series of bursts in cortical nucleus HVC is observed to be tightly locked to song timing (Hahnloser et al. 2002; Lynch et al. 2016).  Control of the timing has been assigned to HVC (Simpson \& Vicario 1990, Ashmore et al. 2005), and to recurrent feedback from the brainstem (McLean et al. 2013, Reinke \& Wild 1998, Striedter \& Vu 1998).  Previous modeling of HVC has invoked a chain-like mechanism to drive downstream areas.  These chain models are instructive but lack biophysical justification.  Meanwhile, models of connectivity downstream, including feedback to HVC, omit important observations of both electrophysiology and of song-related neuronal and respiratory activity.  In particular, the connectivity between HVC and the robust nucleus of the arcopallium (RA) - a connectivity that is critical for song production - is extremely poorly characterized, even though these two regions have received considerable attention by the experimental community.

Here we build upon a biophysically-motivated pattern-generating mechanism in HVC, which has been set forth in a previous paper (Armstrong \& Abarbanel 2016).  We expand on this model to generate representations of acoustic output of the zebra finch, via a specific functional connectivity for downstream regions that is in keeping with observations to date.  

Previous models of HVC have focused on producing the observed sparse bursting of neurons projecting to RA ($HVC_{RA}$ PNs) (Hahnloser et al. 2002; Lynch et al. 2016).  Those models invoke a feedforward chain of excitation (Li \& Greenside 2006; Long et al. 2010; Gibb et al. 2009a, Cannon et al. 2015).  Gibb et al. (2009a) introduced a chain modulated by inhibition, to incorporate evidence that inhibition is integral to the series propagation.  Their proposed mechanism, however, was engineered without biophysical motivations.  Moreover, the chain model is troublesome in that, by its very definition, it does not represent an interconnected web - the picture that emerges from evidence for highly reciprocal structured excitation and structured inhibition within HVC (Kosche et al. 2015).  

Armstrong \& Abarbanel (2016) proposed an alternative to the HVC chain model, in terms of a competition among inhibitory neurons (e.g. Verduzco-Flores et al. 2012; Yildiz \& Kiebel 2011).  This formalism is based on the biophysical process of mutual inhibition.  It readily offers a structured role for inhibition, and permits a formulation that is simpler yet more versatile: a single architecture capable of generating multiple modes of activity.  In this paper, we take that model to drive signaling through a functional architecture of the song motor pathway.

In this paper we focus exclusively on the song motor pathway, and will not discuss its known connections to the auditory system (Vates et al. 1996, Lewandowski et al. 2013) or the anterior forebrain pathway, a circuit that is required for song learning (e.g. Brainard \& Doupe 2002).  The current understanding of the song motor pathway goes as follows.  An initiating signal reaches HVC, and perhaps also the respiratory-related brainstem.  $HVC_{RA}$ PNs then enact a sparse pattern of bursting.  RA, a nucleus long implicated in song generation (Nottebohm et al. 1976, Vu et al. 1994, Yu \& Margoliash 1996, Kubota and Saito 1991, Spiro et al. 1999, Margoliash 1997, Simpson and Vicario 1990), converts the bursts from HVC into more elaborate instructions (Yu \& Margoliash 1996, Margoliash 1997) to be delivered to regions in the brainstem that control respiration and the syrinx, the avian vocal instrument.  The song consists of syllables and inter-syllable gaps, which coincide with active expiration and mini-breaths, respectively. 

Previous modeling aimed to link HVC to song output has been performed (Abarbanel et al. 2004).  The prediction of this model, however, is inconsistent with subsequent observed correlations between RA activity and acoustic structure (Leonardo \& Fee 2005).  In Abarbanel et al. (2004), the syringeal- and respiratory-related brainstem regions were activated sequentially.  This assignment predicted that the number of RA projection neurons (RA PNs) firing should depend on note frequency, and on whether a particular temporal instance occurred during sound versus gap.  In contrast, Leonardo \& Fee (2005) found the number of active RA PNs to be roughly invariant throughout song.  In this paper we show how \textit{simultaneous} signaling by RA to these regions can reproduce that observation.

Previous modeling of a functional feedback loop for song generation has been proposed (Gibb et al. 2009b).  Those authors, however, used the chain model of Gibb et al. (2009a) to describe HVC, and they did not extend the model to acoustic output.  Their schematic for feedback connectivity, on the other hand, is consistent with observed timings of syllables versus gaps (Glaze \& Troyer 2006), and with subsequent air sac pressure timings during song (Andalman et al. 2011).  We build upon this aspect of their framework.  Further, we expand the RA model to include both excitatory and inhibitory populations, and observations that $HVC_{RA}$ PNs excite only the latter (e.g. Spiro et al. 1999).

The work presented in this paper was incited by the question: Can we create a functional connectivity of the song motor pathway that is consistent with observations, such that the HVC model of Armstrong \& Abarbanel (2016) drives a simple representation of song?  We offer an answer, by invoking three features: 1) a detailed functional connectivity between HVC and RA, given known electrophysiology; 2) a specific temporal relationship among signals sent from RA to song-related brainstem regions; 3) song timing that is shared by HVC and recurrent brainstem feedback at the onset of each gap.  The model makes testable predictions regarding the electrophysiology of RA, and of song-related neuronal activity throughout the motor pathway.

\section{MODEL}
\subsection{\textbf{Scope}}

The model, whose scope is shown in Figure 1, is summarized in this Subsection and in Subsection B.  For the interested reader, details of the model are explained in Subsections C - J.  Finally, control of song timing is described in Subsection K.   
\begin{figure}[H]
  \centering
\includegraphics[width=0.4\textwidth]{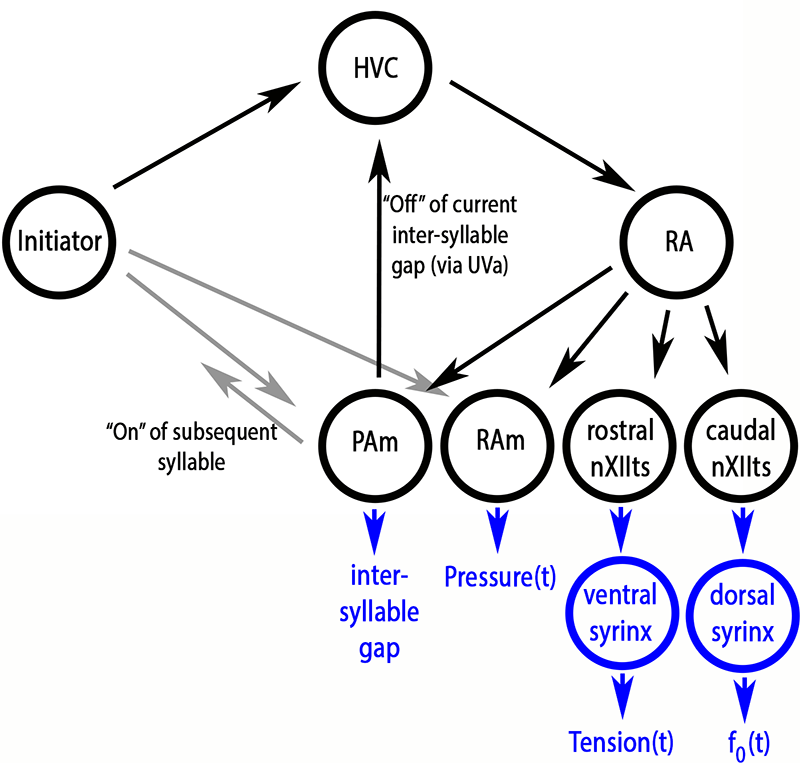}
\caption{\textit{Black and grey}: detailed computational model.  \textit{Blue}: simple one-to-one relations to generate driving forces for song.}
\end{figure}
The model invokes an \lq\lq initiating region\rq\rq\ that is capable of activating HVC via some neuromodulatory mechanism.  We do not explicitly model this initiating region, but rather offer suggestions for its likely geographic location.  The model explicitly includes HVC, RA, and four distinct brainstem regions that have identified roles in driving song.  

These four brainstem regions are: 1) the expiratory-related retroambigualis (RAm), to which is attributed the control of pressure in the air sacs that compress the lungs; 2) the inspiratory-related parambigualis (PAm) (Wild 1993a, Wild 1993b, Wild 1997, Wild et al. 1997, Wild et al. 1998, Roberts et al. 2008); 3) the rostral and 4) caudal tracheosyringeal region of the hypoglossal nerve (nXIIts), which respectively control labial tension (via the ventral syringeal muscle vS) and labial adduction (via a dorsal syringeal muscle that varies across birds) (Wild 1997, Wild 1993b, Vicario 1991a, Vicario 1991b, Gardner et al. 2001, Goller \& Suthers 1996, Larsen \& Goller 2002, Sober et al. 2008).  At the onset of a gap, PAm sends feedback to HVC to halt the currently-active series of $HVC{RA}$ bursts.

Now, in a more realistic picture, these four brainstem regions inter-connect and may overlap.   For simplicity, however, in the zebra finch model we take them to be distinct and non-interacting.  Further, we omit brainstem nuclei whose role in song production appears significant but is to-date obscure\footnote{We omit nucleus DM.  In oscine birds, stimulation to DM elicits calls but not song, so it is generally believed that DM is associated with unlearned vocalization (Wild 1997).}, and we omit audition\footnote{Bottjer \& Arnold (1984) found that adult song is stable after deafening.}.  

\subsection{\textbf{Basic functionality}}

The basic steps for modeling the song motor pathway are as follows.
\begin{enumerate}
  \item During quiescence immediately preceding song, $HVC_{RA}$ neurons are silent above threshold while HVC interneurons are densely spiking, and the opposite situation occurs in RA (e.g. Spiro et al. 1999).  These scenarios are produced by assigning relative threshold potentials for spiking, for the excitatory and inhibitory neurons in each nucleus, respectively.  
  \item Song is initiated via a neuromodulatory mechanism that rapidly increases the strengths of the interneuron-interneuron synapses in HVC\footnote{The initiating signal may target PAm as well, as suggested by Amador et al. 2013; Alonso et al. 2015; Alonso et al. 2016, but that consideration does not affect this model.}, such that a competition is effected among that population.  
  \item Then, due to a specific connectivity within HVC, this competition among interneurons effects a sequence of activations of the $HVC_{RA}$ PNs. 
  \item $HVC_{RA}$ PNs synapse exclusively onto RA interneurons, which are inactive until excited by a projection from HVC (Spiro et al. 1999).  In this model, each $HVC_{RA}$ PN synapses onto four RA interneurons.
  \item Then, due to a specific connectivity within RA, these four RA interneurons suppress a fraction of the (otherwise active) RA PNs.  It is in this way that each $HVC_{RA}$ PN indirectly recruits an ensemble of RA PNs.  
  \item Each of the four RA PNs in an ensemble activates a premotor neuron in a distinct song-related brainstem region. 
  \item Using simple one-to-one rules for brainstem-to-motor connectivity, we offer an informal illustration of how the computational model may reproduce the driving forces for song.  Here, within each $\sim$ 10-ms timebin during a syllable and at the onset of a gap, the four brainstem regions command four distinct motor regions to effect a specific value of: 1) labial tension, 2) air sac pressure, 3) degree of syringeal adduction, and 4) a specific instruction to PAm, respectively.  
  
If the instructions to the four brainstem regions occur simultaneously, then RA PN activity is uncorrelated with the fundamental frequency of the note and whether a syllable or gap is currently playing, as found by Leonardo \& Fee (2005).  RA PN activity is also uncorrelated with note amplitude.
  \item At the onset of an inter-syllable gap, PAm is directed to begin inspiration and send electrical feedback to HVC to silence the currently-active $HVC_{RA}$ PN series.  
  
Throughout the gap, $HVC_{RA}$ PNs continue to fire, until the feedback signal reaches HVC and terminates that series.  During the gap, the four brainstem regions identified above are not \textit{necessarily} activated with the precision required during syllables and gap onsets. 
\end{enumerate}

In Subsections C - J, we describe in detail the mechanics that effect the summary described above.  The uninterested reader may move directly to Subsection K: \textit{Control of Song Timing}. 

\subsection{\textbf{The equations of motion}}

\subsubsection{Neurons}

The neuronal populations are as follows.  HVC contains 12 inhibitory interneurons and 12 $HVC_{RA}$ PNs, distinguished by their leak and synaptic reversal potentials.  RA contains 12 inhibitory interneurons and 12 excitatory PNs that connect to the brainstem.  Again, the inhibitory-versus-excitatory nature of an RA neuron is set by leak and synaptic reversal potentials.  The brainstem consists of four neurons: one representing each of the four regions shown in Figure 1.  Given the dearth of studies on electrophysiology, neuromodulation, and receptor dynamics in the region downstream of RA, we are agnostic regarding whether the brainstem neurons are motor neurons, or excitatory or inhibitory pre-motor neurons.

The electrophysiology of HVC neurons has been studied in detail (Daou et al. 2013).  For simplicity and computational efficiency, however, for all neurons we choose a three-dimensional Hodgkin-Huxley model, as three dimensions is the fewest required to produce bursting (Izhikevich 2007).  The model consists of one compartment with a persistent sodium current and instantaneous activating gating variable, and a fast and a slow potassium current.   

All neurons are three-dimensional Hodgkin-Huxley-type models of the following form:  
\begin{align*} 
  C\diff{V_{i}(t)}{t} &= I_{L,i}(t) + I_{NaP,i}(t) + I_{K_f,i}(t) + I_{K_s,i}(t) 
\end{align*}
\begin{align}
        &+ \sum_{j \neq i}I_{syn_{ij}}(t) + I_{background} + \eta(t).
\end{align}
Here, $V_i$ is the membrane potential of cell $i$, $C$ is the membrane capacitance.  $I_L$, $I_{NaP}$, $I_{K_f}$, and $I_{K_s}$ are ion channel currents: leak, persistent sodium, and fast and slow potassium.  The $I_{syn}$ are the synaptic input currents.  $I_{background}$ represents ambient background excitation, and $\eta$ is a low-noise term.  The neurons are rendered distinct via slightly different values of ion channel maximum conductances.  

The forms of the ion channel currents are:
\begin{align*} 
  I_{L,i}(t) &= g_{L}(E_{L} - V_i(t))\\ 
  I_{NaP,i}(t) &= g_{Na,i} m_{inf,i} (E_{Na} - V_i(t)) \\
  I_{K_f,i}(t) &= g_{K_f,i} n_{f,i}(t) (E_K - V_i(t))\\
  I_{K_s,i}(t) &= g_{K_s,i} n_{s,i}(t) (E_K - V_i(t))
\end{align*}
\noindent
The parameters $g$ and $E$ are the maximum conductances and reversal potentials for each current, respectively.  The neurons are rendered distinct via slightly different values of certain parameters.  One important difference between parameter values of HVC and RA neurons is in the relative excitability of the excitatory versus inhibitory population within each nucleus, described in \textit{Model}.  For the dynamics of fast and slow sodium gating variable $n_f$ and $n_s$, and for all parameter values, see Appendix C.  

The significant differences among the HVC, RA, and brainstem populations are in the synaptic input currents $I_{syn}$.  HVC receives synaptic input from within HVC (and feedback from the brainstem, to be considered below in this Section).  RA neurons receive synaptic input from within RA and from HVC.  Brainstem neurons receive synaptic input only from RA neurons.  Finally, the $I_{background}$ term differs across the three populations by one per-cent (see Appendix C).

\subsubsection{Synapses}

The neurons are connected with chemical synapses, which are modified versions of the form by Destexhe et al. (1994) and Destexhe \& Sejnowski (2001):
\begin{align} 
  I_{syn,ij} &= g_{ij}(T_{max}(t))s_{ij}(t)(E_{syn,i} - V_{i}(t)).
\end{align}
\noindent
Here, $g_{ij}$ are the maximum conductance of a synapse entering cell $i$ from cell $j$.  For HVC interneurons, the $g_{ij}$ are functions of $T_{max}$, the maximum concentration of the relevant neurotransmitter in the vicinity of the synaptic cleft (which, in the avian brain, is $GABA_A$).  $T_{max}$ itself may be time-varying.  (The $g_{ij}$-$T_{max}$ relation will be described below in this Section.)    For all other synapses in this model, the $g_{ij}$ are taken to be static.  The $s_{ij}$ are gating variables, and $E_{syn,i}$ is the reversal potential of cell $i$.  

The ratios of connectivity strengths within HVC, within RA, from HVC to RA, and from RA to Brainstem are roughly 1:10:10:10, respectively.  For the dynamics of gating variables $s_{ij}$, and for all other parameter values, see Appendix C.

These equations were integrated forward to yield voltage traces, via Python's odeINT: a fourth-order adaptive Runge-Kutta scheme.  The time step used was 0.1 ms.

\subsection{HVC functional architecture}

Figure 2 shows the functional architecture of HVC that was set forth in Armstrong \& Abarbanel (2016).  In that paper, we referred to the structure as a functional syllable unit (FSU).  Here, we expand the structure's role to include the onset of a gap, and to in-part encode gap duration.  Thus, in this paper we refer to this structure as a functional HVC unit (FHU). 

In an FHU, three interneurons (triangles: cells numbered 0, 1, and 2) are connected all-to-all, and each interneuron synapses directly to two of three HVC RA PNs (circles: cells numbered 3, 4, and 5)\footnote{Feedback from the excitatory cells is also required for FHU functionality but is not an important consideration in this paper.}.  This structure is capable of displaying multiple modes of activity, depending on the connection strengths among the interneurons $g_{ij}$.  (Hereafter, \lq\lq $g_{ij}$\rq\rq\ refers to coupling strengths among the \textit{interneurons in HVC}.  All other synaptic strengths are taken to be static).  

The two functional modes of interest in this paper are: \lq\lq quiescent\rq\rq\ and \lq\lq active\rq\rq.  Quiescent mode occurs for sufficiently low values of $g_{ij}$.  In this mode, all interneurons may fire simultaneously, thereby suppressing all PNs (not shown).  For higher values of $g_{ij}$, the interneurons may fire in a sequence, thereby effecting sequential firings of each $HVC_{RA}$ PN (Figure 3); we refer to this mode as active.  The toggling between modes we attribute to a neuromodulatory mechanism, to be
\begin{figure}[H]
  \centering
\includegraphics[width=0.4\textwidth]{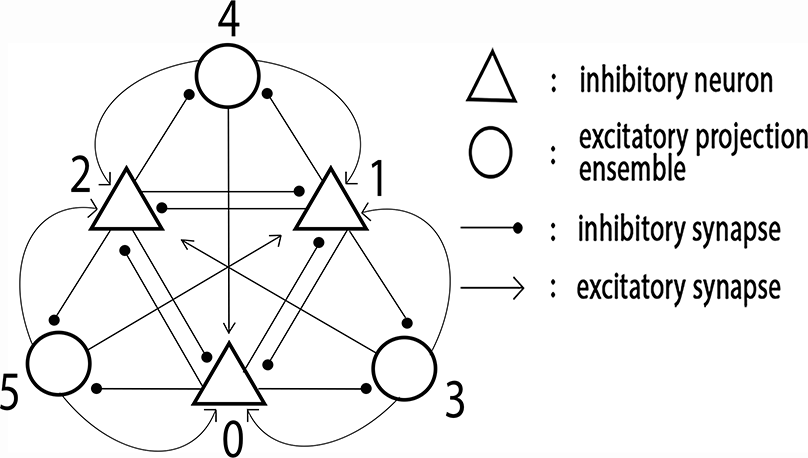}
\caption{A functional HVC unit (FHU).  It is distinct from the usage in Armstrong \& Abarbanel (2016) in that it represents the timing of both a syllable and the onset of the subsequent inter-syllable gap.}
\end{figure}
\begin{figure}[H]
  \centering
\includegraphics[width=0.4\textwidth]{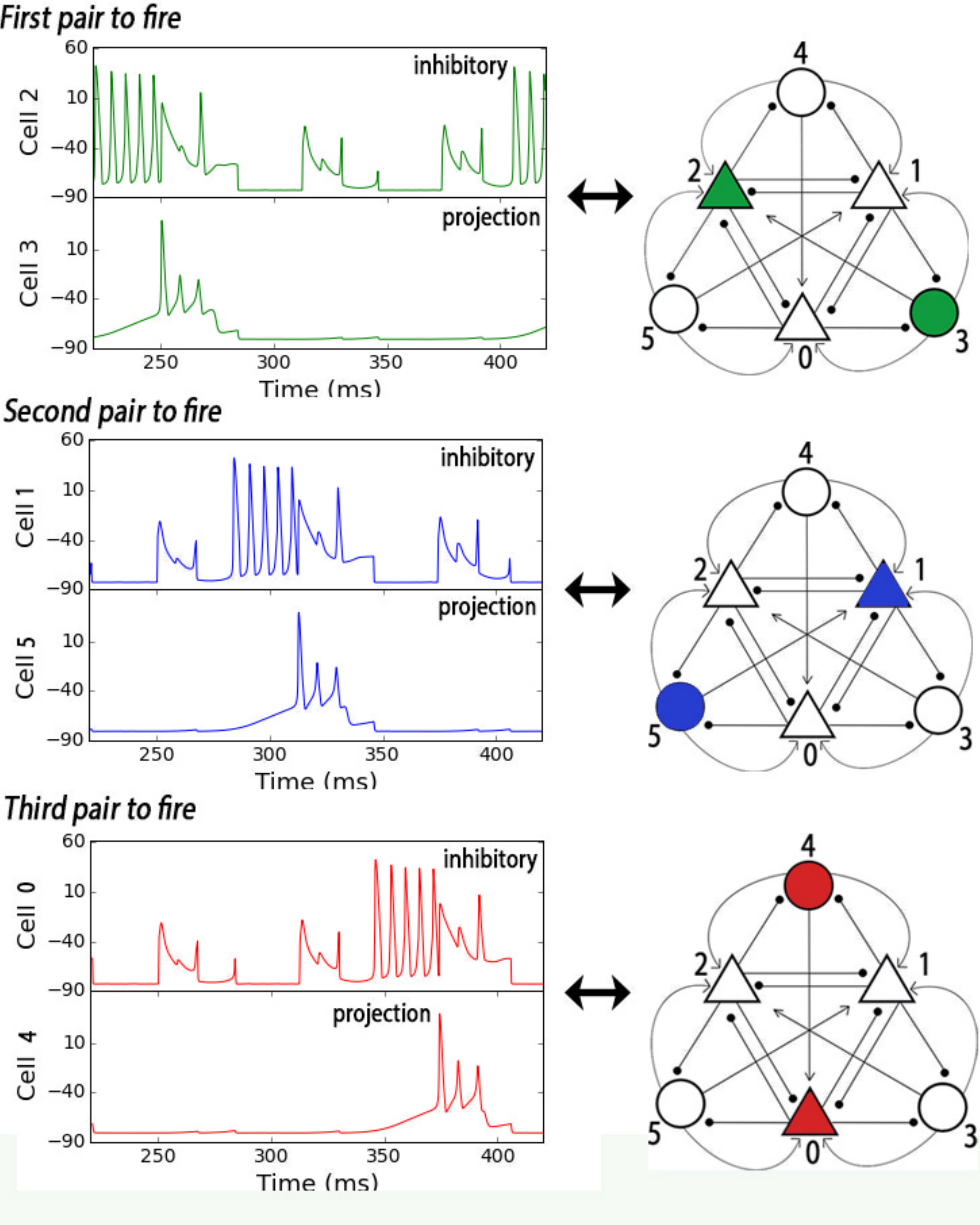}
\caption{A three-frame movie - in green, blue, and red - representing active mode of a functional HVC unit (FHU).  For a certain range of coupling strengths $g_{ij}$ among the interneurons, the interneurons may engage in a series of activations - each of which selects a particular $HVC_{RA}$ PN.  It is in this way that the sparse bursting observed by Hahnloser et al. (2002) and Kozhevnikov \& Fee (2007) may be mimicked.  (\textit{Reproduced from Armstrong \& Abarbanel (2016)}.)}
\end{figure}
\noindent
discussed in Subsection F.

Note that in this paper, each syllable-gap pair is assigned a distinct FHU.  That is: for a motif comprised of four syllables (and three gaps), we take HVC to be comprised of four FHUs.  We are open, however, to an alternative framework in which there exists \textit{one} architecture in all of HVC, which sequentially assumes four distinct functional configurations.  

We find the relative simplicity of the latter framework more appealing.  Meanwhile, we have found the former framework to be simpler to model and describe.  These two distinct frameworks are in agreement, however, on what is essentially being represented in HVC: a sequence of timings of syllable-and-gap pairs, where the identities of syllables and gaps will be encoded downstream, and where the pairs are relatively independent of each other at the level of HVC.  Thus, for the purposes of this paper, we will adopt the former framework.

Finally, for a zebra finch model we assign zero connectivity among these FHUs.  This requirement aims to
\end{multicols}
\begin{figure}[H]
  \centering
\includegraphics[width=0.9\textwidth]{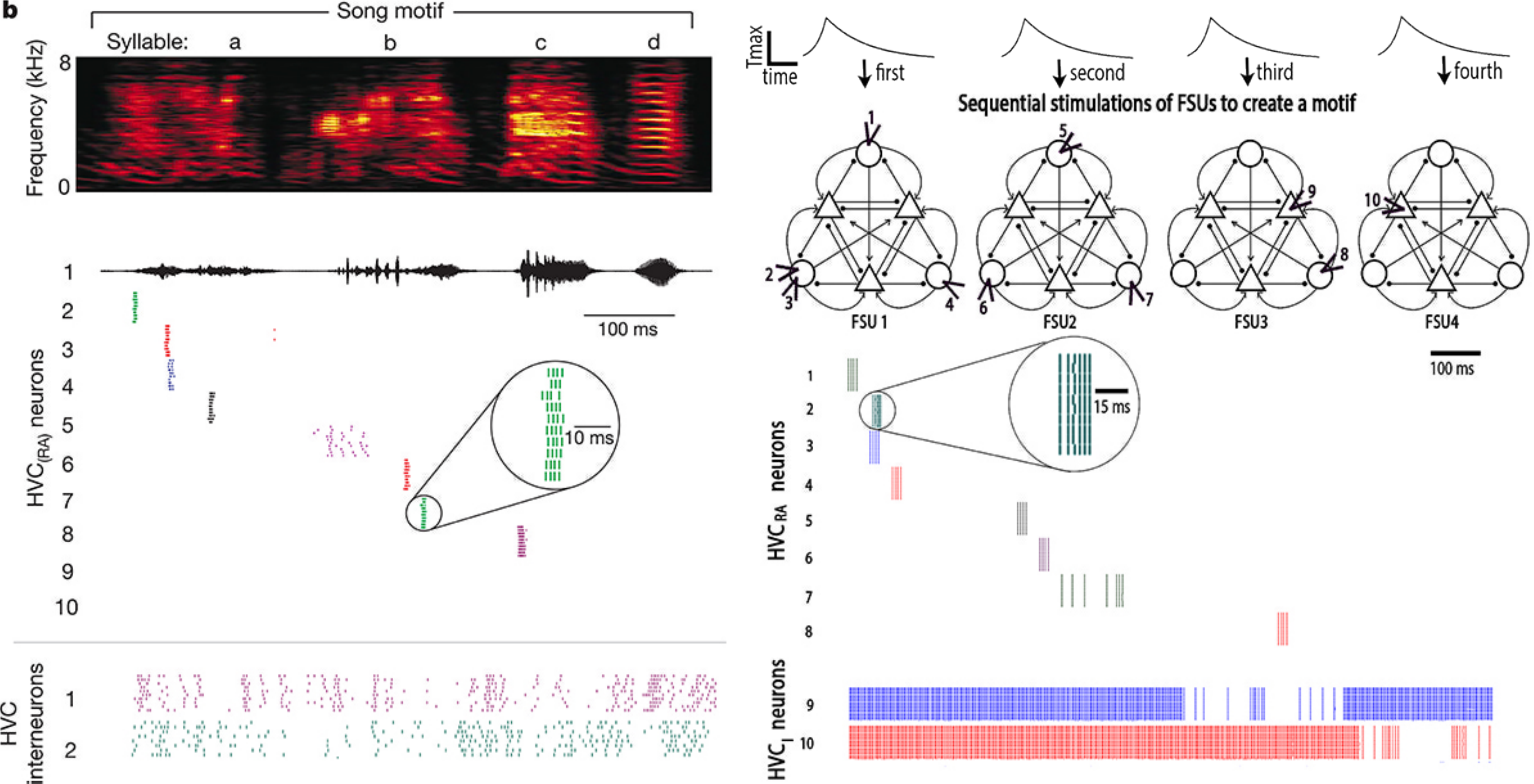}
\caption{Left: Raster plot of $HVC_{RA}$ PNs and $HVC$ interneurons during song (Hahnloser et al. 2002).  Right: Simulated raster plot, using HVC model of Armstrong \& Abarbanel (2016).  Note sparse bursting of $HVC_{RA}$ PNs and dense tonic spiking - with intermittent pauses - of HVC interneurons.  (\textit{Reproduced from Armstrong \& Abarbanel (2016)}).}
\end{figure}
\begin{multicols}{2}
\noindent
mimic observations that individual song syllables \lq\lq freeze out\rq\rq; that is: individual neurons are observed to become locked to particular syllables as song is learned (Vallentin et al. 2016), as well as the prediction by Fee et al. (2004) that learning is most efficient when stages in HVC are decoupled.

\subsection{\textbf{Quiescence immediately prior to song}}

Immediately prior to song, interneurons in HVC are spiking densely, and $HVC_{RA}$ PNs are silent above threshold.  This scenario is captured by setting the leak reversal potential for interneurons slightly lower than that for $HVC_{RA}$ PNs, and providing the circuit with a low-amplitude background current.

Meanwhile, the opposite scenario occurs in RA: interneurons are silent, while the RA PN population is active (Spiro et al. 1999).  This scenario was reproduced by setting the leak current reversal potential slightly lower for the RA PNs compared to the value for RA interneurons.

\subsection{\textbf{Initiation: neurmodulation alters couplings of HVC interneurons to effect patterned activity.}}

At song onset, the FHU corresponding to the first syllable-gap pair is initialized to active mode.  The initiator we attribute to a neuromodulatory mechanism that is capable of rapidly increasing the interneuron-interneuron coupling strengths $g_{ij}$.  We do not describe this mechanism, other than suggesting that it originates in the brainstem\footnote{Alonso et al. (2016), in a population model of song production, assumed the initiator to be in the brainstem, for a different reason: non-songbirds lack the telencephalic structures but can produce song-like sounds.}.  The brainstem is a likely location given that it contains neuromodulatory neurons that make long-range dopaminergic projections to the telencephalon\footnote{These dopaminergic brainstem regions are: substantia nigra pars compacta (SNc) and VTA (Gale \& Perkel 2006).}, and that a direct connection from the ventral tegmental area (VTA) to HVC has been identified (Hamaguchi \& Mooney 2012).  

Equation 2 formalizes the mechanism by which neuromodulation alters the coupling strengths $g_{ij}$.  For inhibitory-inhibitory connections, the $g_{ij}$ are functions of maximum neurotransmitter concentration $T_{max}$.  The function $T_{max}(t)$ is a steep rise and gradual decay, and $g_{ij}$($T_{max}(t)$) goes as $\alpha T_{max}^{\beta}$.

This model reproduces basic qualitative features of HVC interneurons and RA-projecting PNs during song and during quiescence.  Figure 4, reproduced from Armstrong \& Abarbanel (2016), compares a simulated raster plot to that of Hahnloser et al. (2002), where $HVC_{RA}$ PNs burst sparsely while interneurons spike tonically with intermittent pauses, over ten song renditions (this finding by Hahnloser et al. (2002) has been significantly enhanced by Lynch et al. (2016)).  The FHU structure also roughly captures the observed high rates of reciprocal connectivity between HVC interneuron and PN populations, and that inhibition masks the activity of an excitatory population (Kosche et al. 2015).  

\subsection{\textbf{A similar structure is created in RA, whose coupling strengths are static and where the inhibitory-to-excitatory connections are critical for functionality.}}

We base our construction of RA upon three experimental findings: 1) During quiescence, RA PNs are active and RA interneurons are silent above threshold (Spiro et al. 1999).  2) $HVC_{RA}$ PNs synapse to RA interneurons and not to RA excitatory PNs (Kubota \& Saito 1991, Mooney \& Konishi 1991, Spiro et al. 1999).  3) RA interneurons, which are inhibitory, make vast projections to RA PNs (Spiro et al. 1999), which are excitatory.

First we reproduce quiescence in RA by setting the leak reversal potential $E_L$ for the interneurons to be lower that than for the excitatory neurons: -85 versus -80 mV.  Then, when both populations receive the same low background excitation, the PNs burst and the interneurons are inactive above threshold.

Next we create a connectivity in RA such that excita- 
\begin{figure}[H]
  \centering
\includegraphics[width=0.4\textwidth]{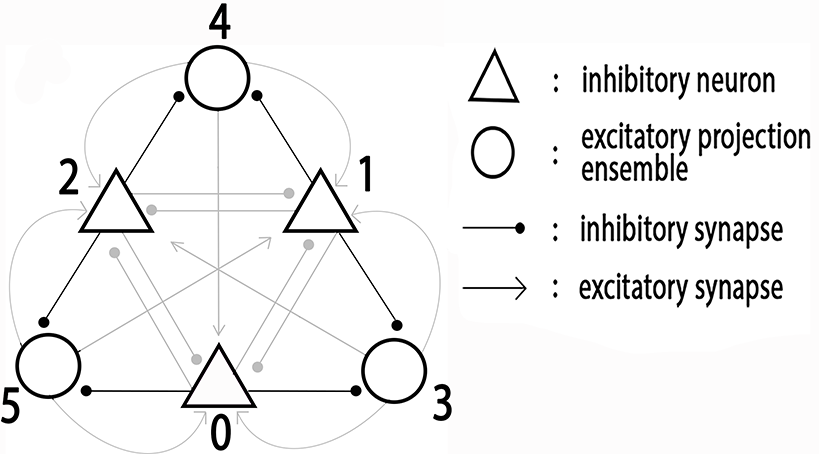}
\caption{One of four identical structures in RA.  These are \textit{not} the FHUs of HVC; the synaptic strengths are static.  The inhibitory-to-inhibitory and excitatory-to-inhibitory connections, while permitted, are cast in grey for de-emphasis.  The only connections critical to the functionality of this model are the inhibitory-to-excitatory connections (black).}
\end{figure}
\noindent
tion of interneurons will suppress a fraction of the excited RA PN population.  Figure 5 shows a schematic: an architecture taken from the FHU in HVC.  Now, we emphasize that this structure in RA is \textit{not} an FHU: the connectivity strengths are static.  Interneuron-to-interneuron and excitatory-to-inhibitory connections are permitted (grey lines on Figure 5), but they are de-emphasized.  The critical connections in RA are the inhibitory-to-excitatory projections (black in Figure 5).  Note that, as in HVC, there are three interneurons (triangles) and three PNs (circles) per structure, and that each inhibitory neuron directly connects to two out of three of the excitatory cells.

Finally, we take RA to be comprised of four such structures.  These structures correspond to distinct geographical regions in RA, each of which is known to make projections to one of the four distinct brainstem regions noted in Subsection A.  Specifically: rostral, ventral, and dorsal RA project to rostral nXIIts (for control of note frequency), caudal nXIIts (for control of adduction), and the respiratory regions (RAm and PAm), respectively.

\subsection{\textbf{An HVC PN locks with an RA PN ensemble indirectly, via direct connections to RA interneurons.}}

Leonardo \& Fee (2005) found that the activity of RA PN ensembles is strikingly similar to the activity of $HVC_{RA}$ PNs during song, in that they occur on a $\sim$ 10-ms timescale and bursts are tightly locked to specific temporal locations during song.  Both slowly-and-rapidly-changing acoustic structure was associated with the fastest timescale of change of RA ensembles (10 ms).  Further, the number of RA PNs per ensemble was essentially invariant over sound versus inter-syllable gaps, and over sounds of differing fundamental frequencies\footnote{Leonardo \& Fee (2005) found that the number of RA PNs firing did correlate with harmonic-versus-non-harmonic sounds.  The zebra finch alternates high-frequency pure tones with low-frequency sounds that are spectrally rich (Sitt et al. 2008), and in this paper, for simplicity, we implicitly ignore the latter.}.  Identical syllables possessed the same ensembles; similar sounds did not.  In addition, Yu \& Margoliash (1996) found that $HVC_{RA}$ PN identity is correlated with syllable identity, while the RA PN identity is more strongly correlated with note identity.  

Taking these findings together, we chose an HVC-to-RA connectivity\footnote{Feedback connections from RA to HVC have been identified (Basista et al. 2014) but are ignored here.} in which each 10-ms burst of an $HVC_{RA}$ PN recruits an RA PN ensemble.  The number of RA PNs per ensemble should be relatively invariant at each temporal instance of song, and on a raster plot the identities of RA PNs within each ensemble should \end{multicols}
\begin{figure}[H]
  \centering
\includegraphics[width=0.7\textwidth]{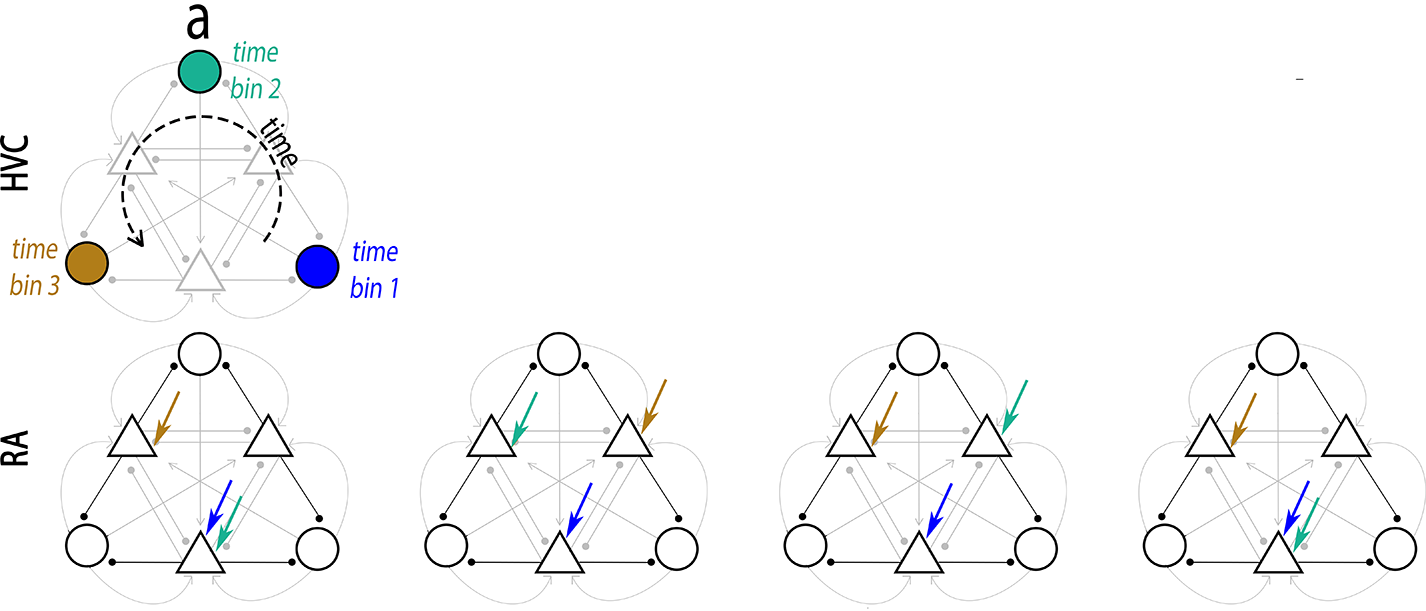}\\
\hrule 
\includegraphics[width=0.7\textwidth]{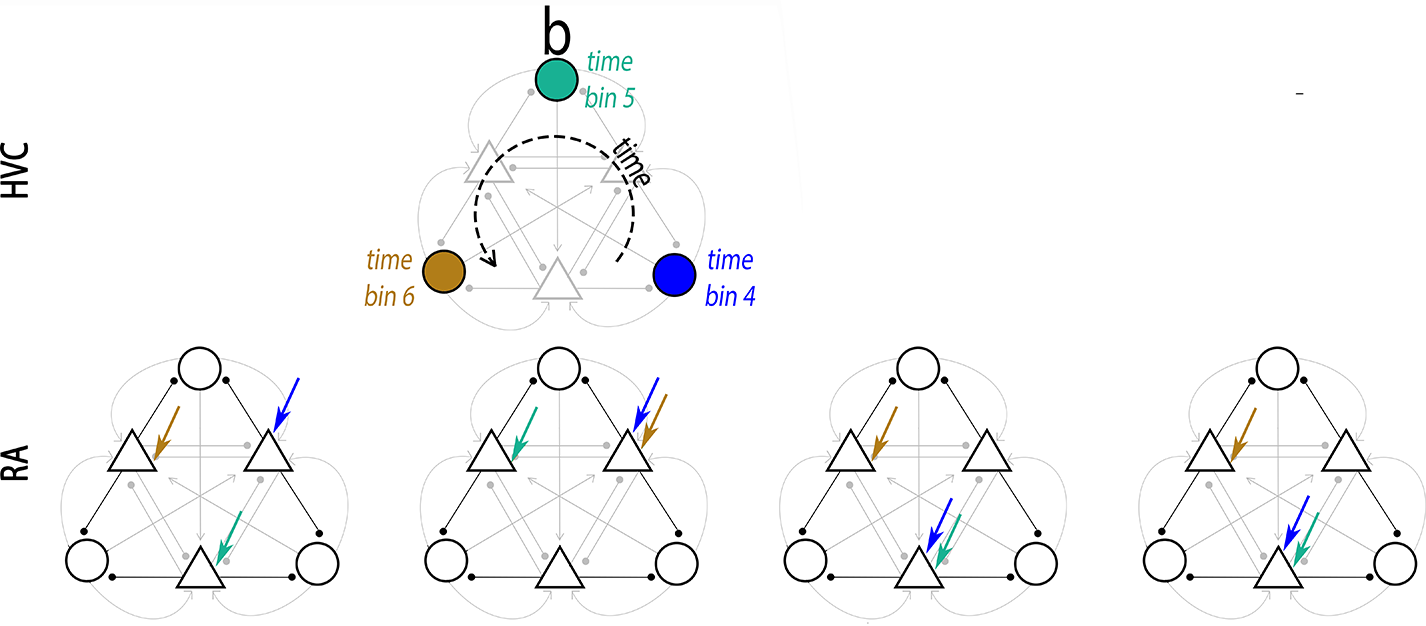}\\
\hrule
\includegraphics[width=0.7\textwidth]{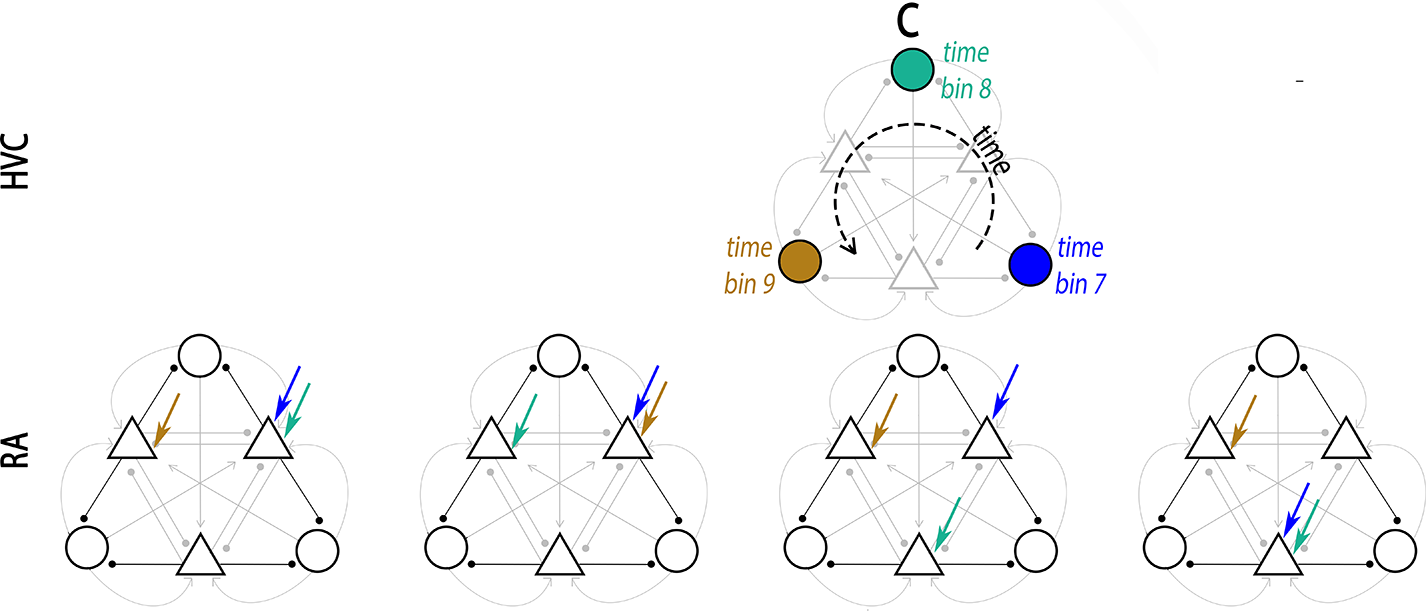}\\
\hrule
\includegraphics[width=0.7\textwidth]{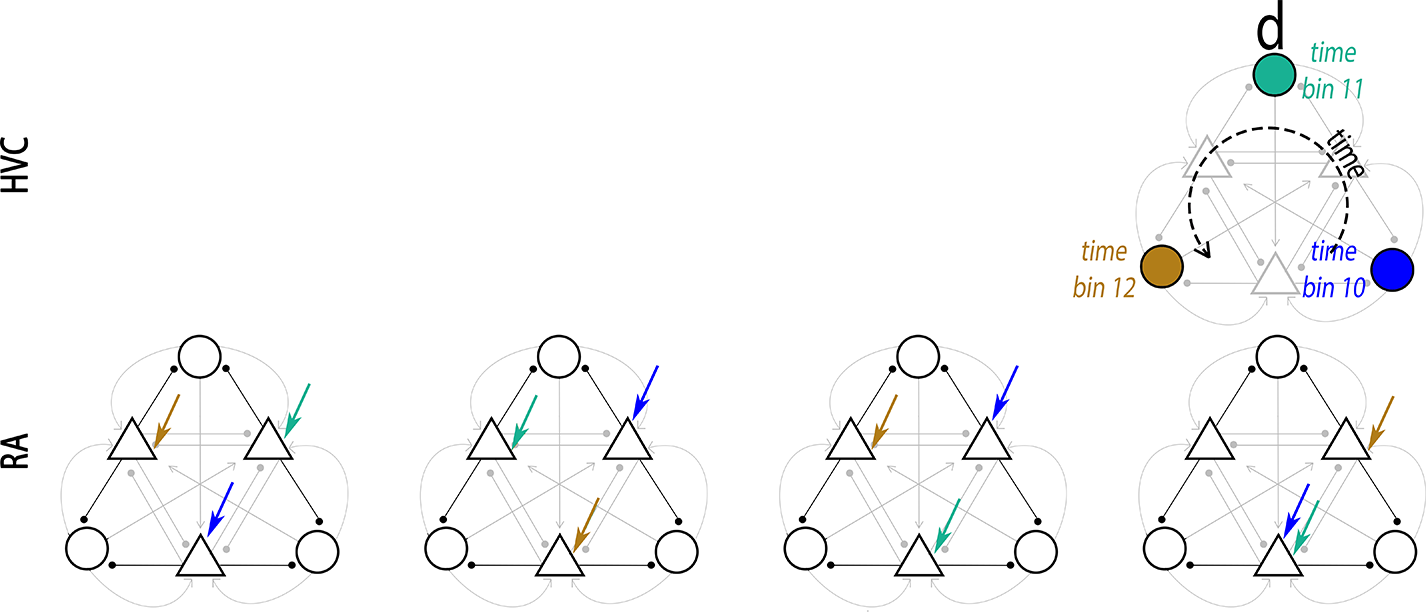} 
\caption{HVC-to-RA connectivity for a full motif, represented by a four-frame movie.  Here, four FSUs in HVC are sequentially activated: one representing syllable $a$, $b$, $c$, and $d$, and the ensuing gap for each.  Each FSU targets the same four structures in RA.  Each $\sim$ 10-ms burst of an $HVC_{RA}$ PN codes for a $\sim$ 10-ms \lq\lq time bin\rq\rq\ within a syllable or at gap onset.  This connectivity, together with the RA-to-brainstem connections of Figure 7, created the raster plot of HVC, RA, and brainstem activity during song shown in \textit{Results}.}
\end{figure}
\begin{figure}[H]
  \centering
\includegraphics[width=0.7\textwidth]{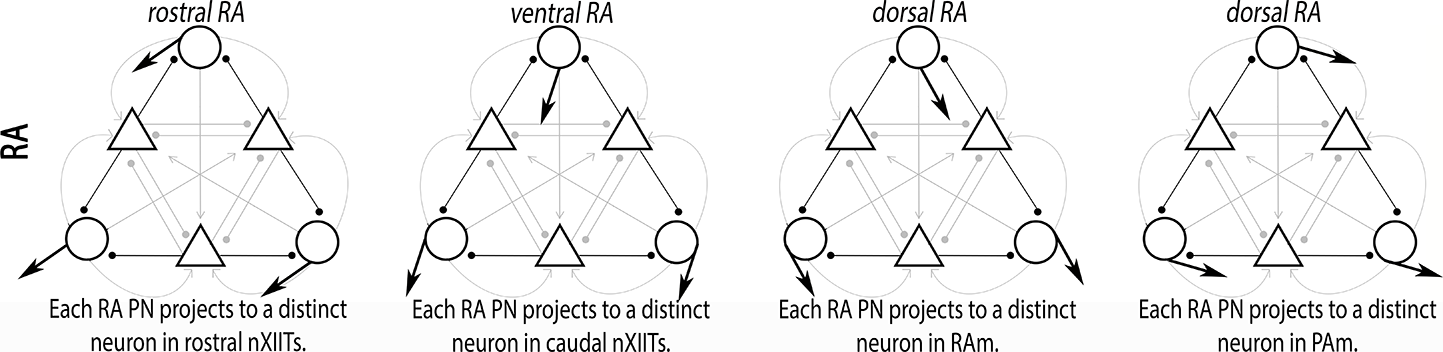}
\caption{RA-to-brainstem connectivity for a full motif.  Each RA PN connects directly (black arrows) to a specific brainstem neuron, which is not pictured.  RA PNs in the first, second, third, and fourth structures of RA project to specific neurons in rostral nXIIts, caudal nXIIts, RAm, and PAm, respectively.  This connectivity, together with the HVC-to-RA connections of Figure 6, created the raster plot of HVC, RA, and brainstem activity shown in \textit{Results}}.
\end{figure}
\begin{multicols}{2}
\noindent
not appear to be correlated with the identities of RA PNs within other ensembles.  

The specifics are as follows.  Each $HVC_{RA}$ PN projects directly to one randomly-selected interneuron in each of the four RA structures.   Then the connectivity of Figure 5 is such that when any one interneuron becomes excited, it suppresses the activity of two of the three\footnote{The observed fraction is ten per cent (Leonardo \& Fee 2005).} PNs in that structure - and permits the third to burst (as is the case within an FHU of HVC).  It is in this way that each $HVC_{RA}$ PN indirectly recruits - via the RA interneuron population - an ensemble of RA PNs.  For an illustration of this mechanism, see Appendix A.

Figure 6 contains the full schematic for HVC-to-RA connectivity.  It consists of four sequential panels, one for each FHU.  For example, the top panel of Figure 6 illustrates the connectivity involving the FHU representing the first syllable-gap pair (\lq\lq a\rq\rq), and the four structures in RA.  Here, a blue arrow entering an interneuron in any of the four structures of RA indicates that the blue $HVC_{RA}$ PN that bursts during the first \lq\lq time bin\rq\rq\ of song directly excites those particular interneurons during that first time bin.

\subsection{\textbf{RA simultaneously signals four distinct brainstem regions, during each time bin of a syllable and at gap onset.}}

Each of the four RA PNs in a currently-active ensemble then directly connects to a neuron in one of four distinct brainstem regions shown in Figure 1.  Figure 7 shows the projections leaving RA for these regions (brainstem regions are not shown).  Specifically, the first, second, third, and fourth RA structures represent rostral, ventral, dorsal, and dorsal RA, respectively.  Signals are sent: 1) from rostral RA to ventral/rostral nXIIts; 2) ventral RA to ventral/caudal nXIIts; 3) dorsal RA to RAm; 4) dorsal RA to PAm.  We choose these regions for their direct connections to labial tension, syringeal adduction, air sac pressure, and inspiration, respectively.  Importantly, in this model these four signals are sent essentially simultaneously and continuously (on a $\sim$ 10-ms timescale) throughout each syllable and gap onset.  

As relatively little is known about neurmodulation and receptor dynamics in the brainstem (e.g. Schmidt \& Wild 2014), we assume a one-to-one relation between each of these four RA structures and each of the four brainstem areas.  That is: Each RA PN effects the excitation of a motor neuron (which in vertebrates are excitatory), either by directly exciting it or indirectly via interactions with pre-motor brainstem neurons.  

Note that by assigning specific brainstem regions to specific geographical locations in RA, we have implicitly assumed no long-range connectivity within RA.  There are known vast connections across RA, particularly between ventral and dorsal regions (e.g. Spiro et al. 1999).  Note also the assumption of no cross-connectivity at the brainstem level.  There exists an extensive literature on dense respiratory-syringeal connections at the brainstem level (e.g. Schmidt \& Wild 2014).  These omissions will be addressed in \textit{Discussion}.  

\subsection{\textbf{Switching among FHUs occurs via feedback to HVC from downstream.}}

Amongst PAm, RAm, and nXIIts, the only robustly identified\footnote{It is unlikely that the syrinx sends feedback to the motor pathway.  Ashmore et al. (2005) found that stimulation to nXIIts distorted sound but did not affect song timing or structure.  Further, learned song can be destroyed via lesions to HVC and RA but is unaffected by syringeal denervation (Simpson \& Vicario 1990; Vicario 1991b; Wild 1997).  RAm appears to not project to Uva (McLean et al. 2013).} feedback pathway to HVC is via Uva (Mooney 2009) from PAm (McLean et al. 2013, Reinke \& Wild 1998, Striedter \& Vu 1998).  In addition, activity in Uva shortly precedes the onset of each syllable (Aronov \& Fee 2008).  In this paper, then, we take feedback to HVC to occur via PAm\footnote{HVC may receive feedback on a timescale faster than a syllable.  The coordination of the HVC hemispheres, for example, occurs on a timescale of 25-50 ms throughout song (Ashmore et al. 2004), and the mechanism effecting that coordination is unknown.  Continual feedback faster than a syllable is not required for the model in this paper.}.  Specifically: at the onset of an inter-syllable gap, PAm receives an order to send feedback to HVC.  

There are two issues here, however, to consider.  First: how does a succession of FHUs become activated?  Second: how does a succession of FHUs become \textit{de}activated?  Now, in Armstrong \& Abarbanel (2016) we took the inactivation of an FHU to occur once the temporarily-enhanced neurotransmitter concentration had decayed below some critical value.  This was not a strong argument, given the 10-ms temporal precision implicitly required of neurotransmitter decay (the required rise time was not a concern, as these can be nearly instantaneous; see references in Armstrong \& Abarbanel; 2016).  In this paper we suggest instead that an FHU is deactivated once electrical feedback reaches HVC from the brainstem.

That is: to coordinate the \lq\lq off\rq\rq\ of the current FHU with the \lq\lq on\rq\rq\ of the subsequent FHU requires that the electrical signal from PAm (via Uva) and the neuromodulatory signal from, for example, VTA, reach HVC essentially simultaneously.  For this reason, we tentatively suggest that the series of signals to the initiator are triggered by PAm as well; see \textit{Discussion}.

Note that in this paper the neuromodulatory \lq\lq on\rq\rq\ signal to HVC is modeled in the $T_{max}$-$g_{ij}$ relation of Equation 1.  To model the electrical \lq\lq off\rq\rq\ signal, we merely truncate the integration.

\subsection{\textbf{Control of timing}}

We base our assignment of song timing on two bodies of observations.  First, the number of RA PNs firing at any instance during song is independent of whether the instance occurs during a syllable or a gap   (Leonardo \& Fee 2005).  If HVC RA PNs are locked to these ensembles, then they should possess the same essential relationship over syllables versus gaps.  Second, Ashmore et al. (2005) found via electrical stimulations to HVC that the time to motif truncation was independent of whether the stimulation was given during a note or a gap.  Third, and perhaps most intriguing, is the finding of those authors that when a syllable truncated but song continued, the gap lengthened so that the total motif duration was preserved.  These lines of evidence indicate that the sequence of HVC RA PNs represents a continuous timing throughout the full motif, regardless of whether a particular time bin is encoded downstream as a syllable or gap.

On the other hand, a second set of observations indicates that the timing mechanism is not entirely dictated by HVC.  Andalman et al. (2011) found that cooling HVC stretches air sac pressure time series differentially.  During syllables and during roughly the second half of inspiration, these time series stretch in response to HVC cooling.  The first half of inspiration, however, remains essentially unchanged.  Further, Glaze \& Troyer (2006) have identified significantly higher variability in gaps versus syllables.   

We attribute this second collection of findings to the timescales of signals returning to HVC, both from PAm and from the source of neuromodulation - where the signal from PAm is triggered \textit{by HVC} at the onset of each inter-syllable gap.  

Specifically, the timing framework is as follows.  The sequence of $HVC_{RA}$ PN bursts encodes syllable duration, including the onset of the subsequent gap.  The gap onset is coded via a signal to PAm to begin inspiration.  The $HVC_{RA}$ PN activity remains self-sustaining until a signal from PAm arrives and terminates it.  A series of syllables is invoked via recurrent activation from the initiating region, which may also be triggered by PAm; see \textit{Discussion}.

Note that in this small model, each 10-ms timebin of song is encoded by one $HVC_{RA}$ PN.  On a larger scale, this by no means need be the case.

\section{RESULTS}
Figure 8\footnote{To create the raster plot, we ran the time series using the $g_{ij}$-$T_{max}$ relation described in Appendix A.  That is: we invoke a neuromodulatory mechanism to commence each syllable.  To represent the electrical feedback ceasing each inter-syllable gap, we simply truncated the time series.} shows a simulated raster plot over five song renditions, based on the connectivity shown in Figures 6 and 7.  The raster plot shows eight $HVC_{RA}$ PNs, nine RA PNs, and two brainstem neurons.  The spectrogram at top results from simple one-to-one assignments between brainstem and motor areas yielding the driving forces \end{multicols}
\begin{figure}[H]
  \centering
\includegraphics[width=0.55\textwidth]{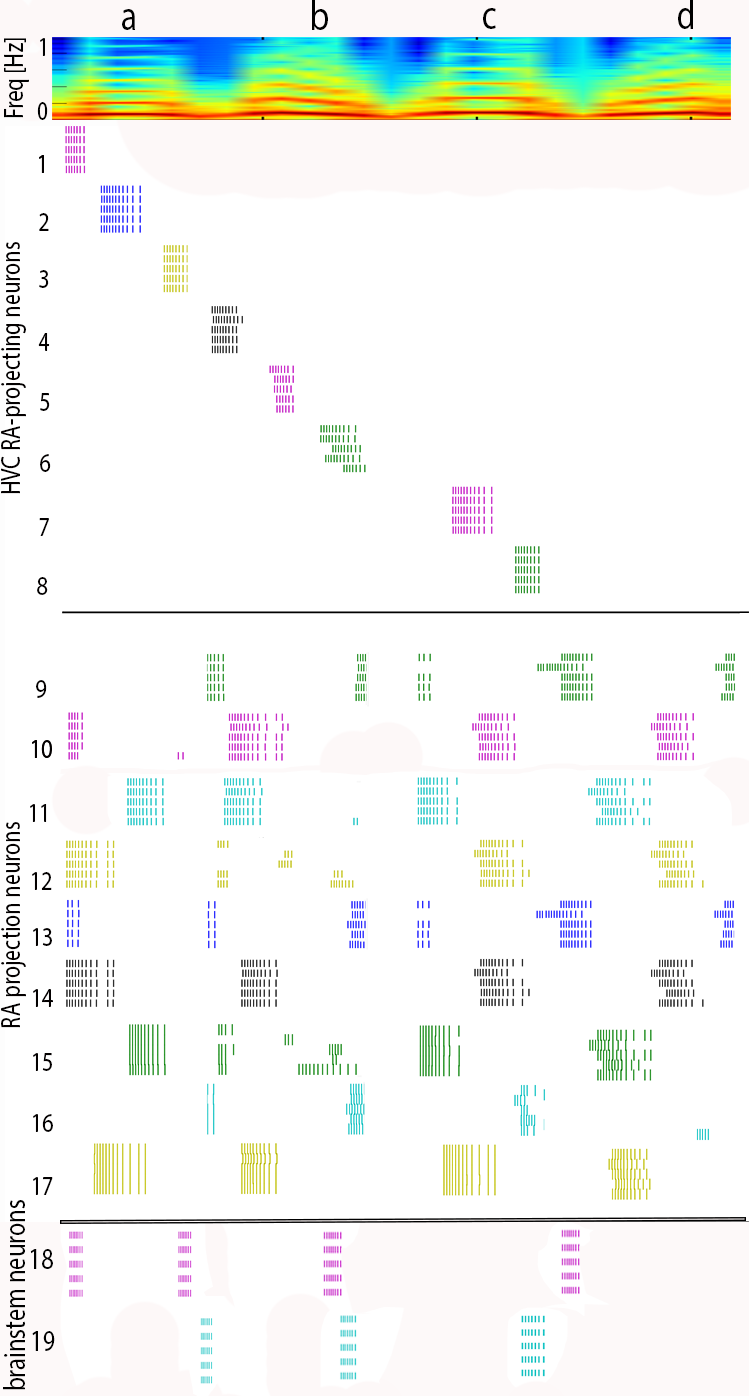} 
\caption{Simulated raster plot of $HVC_{RA}$ PNs and RA PNs over five song renditions, via the connectivity of Figures 6 and 7.  Each $HVC_{RA}$ PN is locked to a particular RA PN ensemble.  Numberings correspond to the numberings of electrodes in Figure 9.  $HVC_{RA}$ firings may be compared to the experimental raster plot of Hahnloser et al. (2002): Figure 4, left panel; RA PN firings may be compared to the experimental raster plot of Figure 10.  Note that the firings of specific brainstem neurons are tightly locked to particular temporal instances during song.  The spectrogram at top was created via simple assignments to motor areas (Appendix B), which generated the driving forces for song (Figure 11, left panel).}
\end{figure}
\begin{figure}[H]
  \centering
\includegraphics[width=0.8\textwidth]{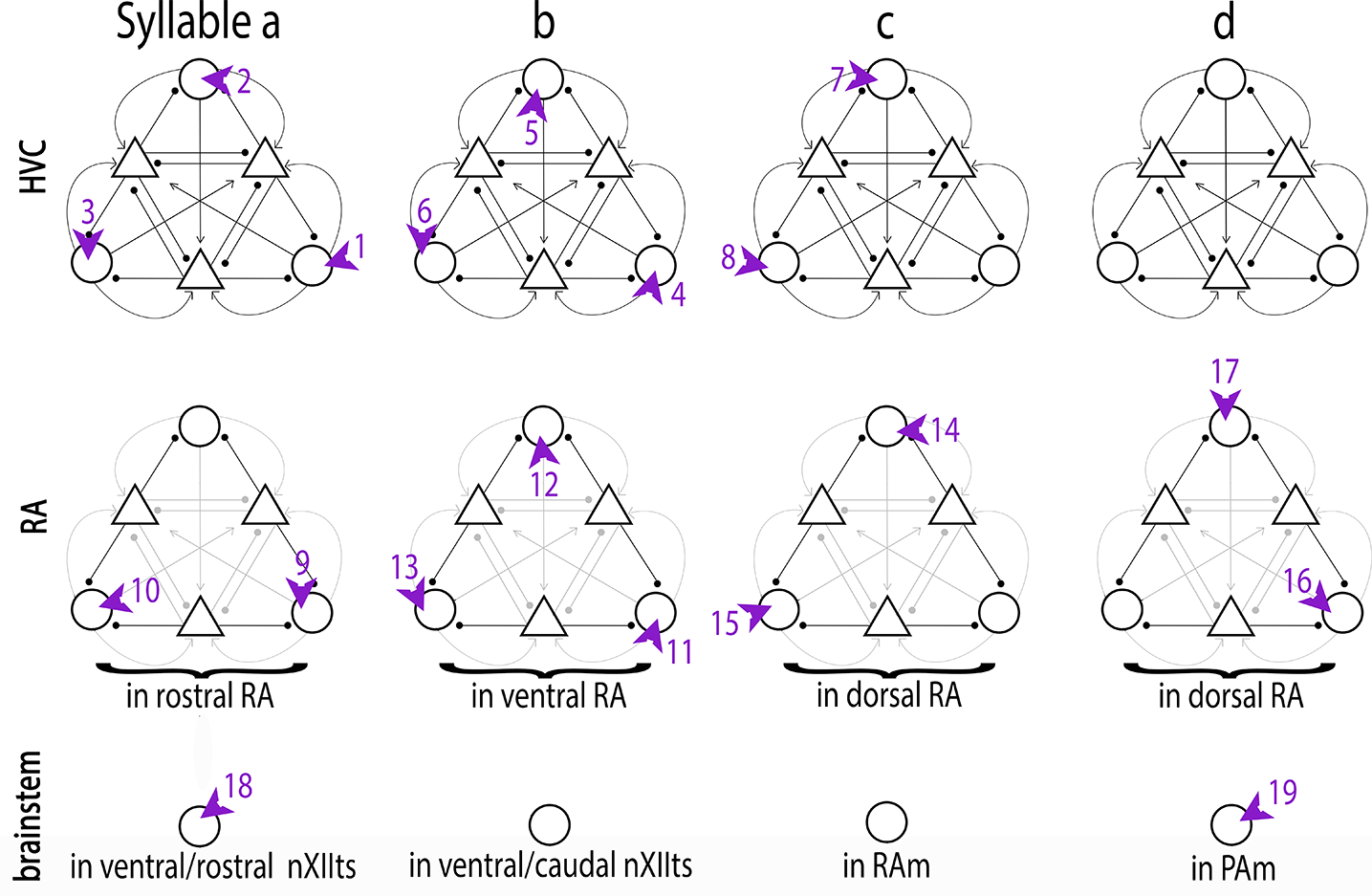} 
\caption{Model schematic with electrodes placed by an experimentor who has inadvertently targeted particular neurons.  Numbers on electrodes correspond to vertical neuron numberings on the raster plot of Figure 8.}
\label{fig:schematicFull}
\end{figure}
\begin{multicols}{2}
\noindent
for song, to be described below in this Section.  

Figure 9 shows a schematic indicating the identities of neurons numbered 1 - 19 on the raster plot.  In Figure 9, electrodes - with corresponding numbers 1 - 19 - have been placed inadvertently by an experimenter into particular neurons in HVC, RA, and the brainstem. 

The reader may compare the raster plot of the HVC neurons to the sparse bursting found by Hahnloser et al. (2002) (Figure 4, left panel), and that of the RA interneurons to Figure 10, reproduced from Leonardo \& Fee (2005).  Note that the bursts of the two brainstem neurons (18 and 19 in Figure 8) indicate that at least a subpopulation of brainstem neurons should be found to be tightly locked to certain temporal instances during song - or, perhaps more tellingly, to certain discontinuities in the time series of driving forces.  See Subsection B for the explicit link between the bursting of the brainstem neurons 18 and 19 and specific instances during the time series of the driving forces.

We note some other features captured by the simulated raster plot, all noted by Leonardo \& Fee (2005).  First, each $HVC_{RA}$ PN is tightly locked with an ensemble of RA PNs upon repeated song renditions.  Second, the activity of each RA PN is similar to that of the HVC PNs in that they burst at reliable temporal locations.  Each RA PN, however, bursts multiple times during each rendition, and in some cases multiple times within one syllable.  Third, there is little variation in the number of RA PNs bursting over temporal locations, including gaps versus syllables.  
\subsection{\textbf{Example driving forces for song}}

Below the brainstem we assigned simple one-to-one relations between each brainstem and particular motor orders.  The resulting time series of driving forces are depicted at left in Figure 11.  See Appendix B for the complete list of instructions used to generate these driving forces. 

In the left panel of Figure 11, the third, fourth, and fifth rows are time series of $b$, $f_0$, and $k$ - the parameters governing labial displacement $x(t)$ (first row):
\begin{align*}
  \dot{x} &= y;
\end{align*}
\begin{align}
  \dot{y} &= -kx - cx^2 + by - f_0
\end{align}
\noindent
(Laje et al. 2002).  In the second line of Equation 3: the first term on the right side represents restitution, where $k$ is a spring constant with roughly a one-to-one correlation with labial tension $T$.  The second term is a
\end{multicols}
\begin{figure}[H]
  \centering
\includegraphics[width=0.9\textwidth]{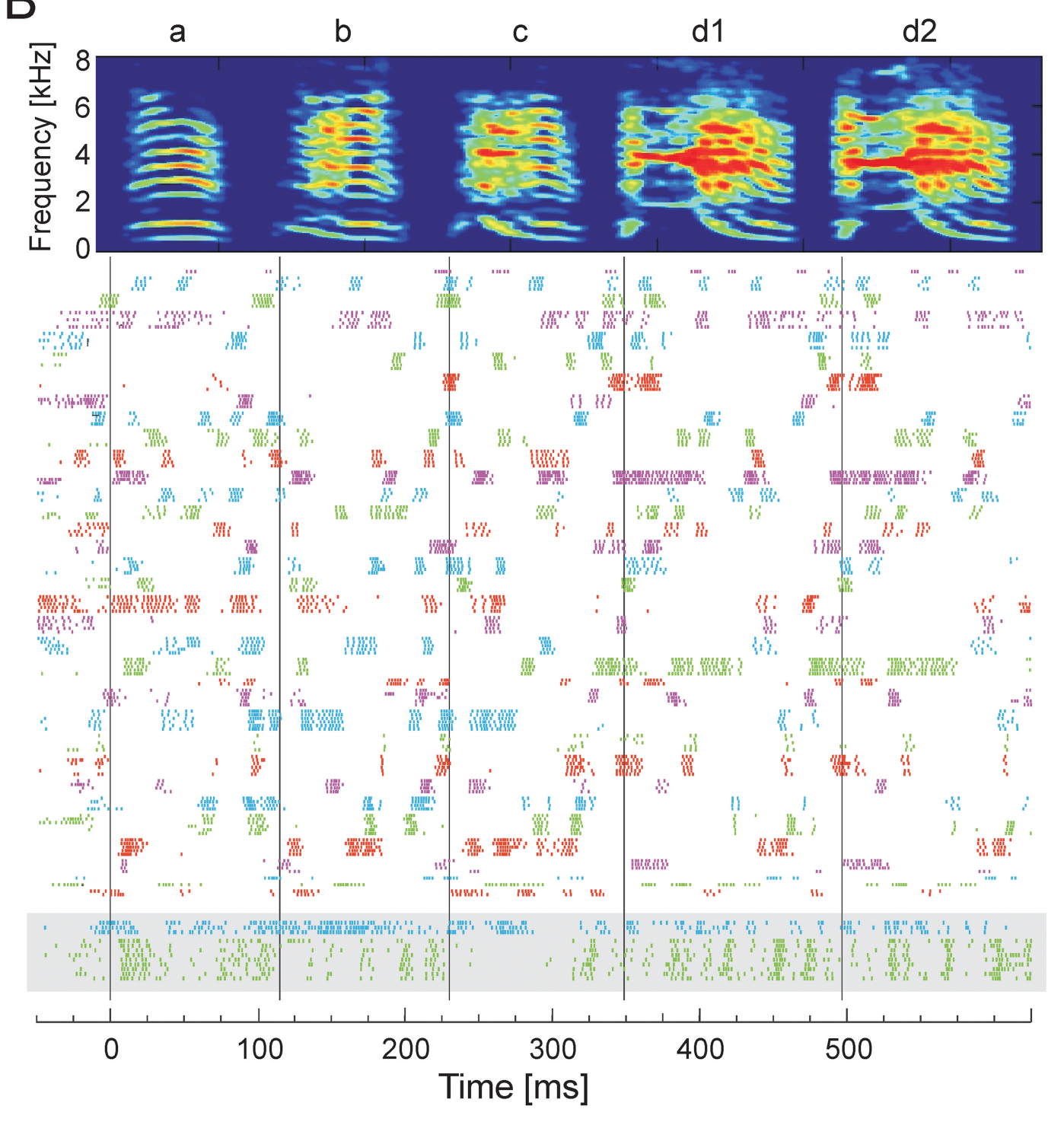} 
\caption{Experimental raster plot of RA PNs during song, for comparison with the simulated raster plot of Figure 8.  (\textit{Reproduced from Leonardo \& Fee (2005).})}
\end{figure}
\begin{multicols}{2}
\noindent
nonlinear dissipation term associated with the labia meeting each other or the containing walls; we take $c = 0.1$.  The term $by$ is a function of the driving air sac pressure $P$; $P$ relates one-to-one to $b$.  Finally, the driving force $f_0$ is set by adduction or abduction of the syrinx\footnote{The muscle governing adduction may vary across birds (Vicario 1990, Larsen \& Goller 2002, Goller \& Suthers 1996), and in this paper we do not attribute the force $f_0$ to a specific muscle.}.

The first row of the left panel in Figure 11 shows the time series of labial displacement $x(t)$.   The second row shows $x(t)P(t)$, which is roughly representative of an acoustic pressure wave (see Appendix B).  The Fourier transform of $x(t)P(t)$ yielded the spectrogram shown in the simulated raster plot of Figure 8.

For comparison, at right in Figure 11 are the time se-
\end{multicols}
\begin{figure}[H]
  \centering
\includegraphics[width=0.4\textwidth]{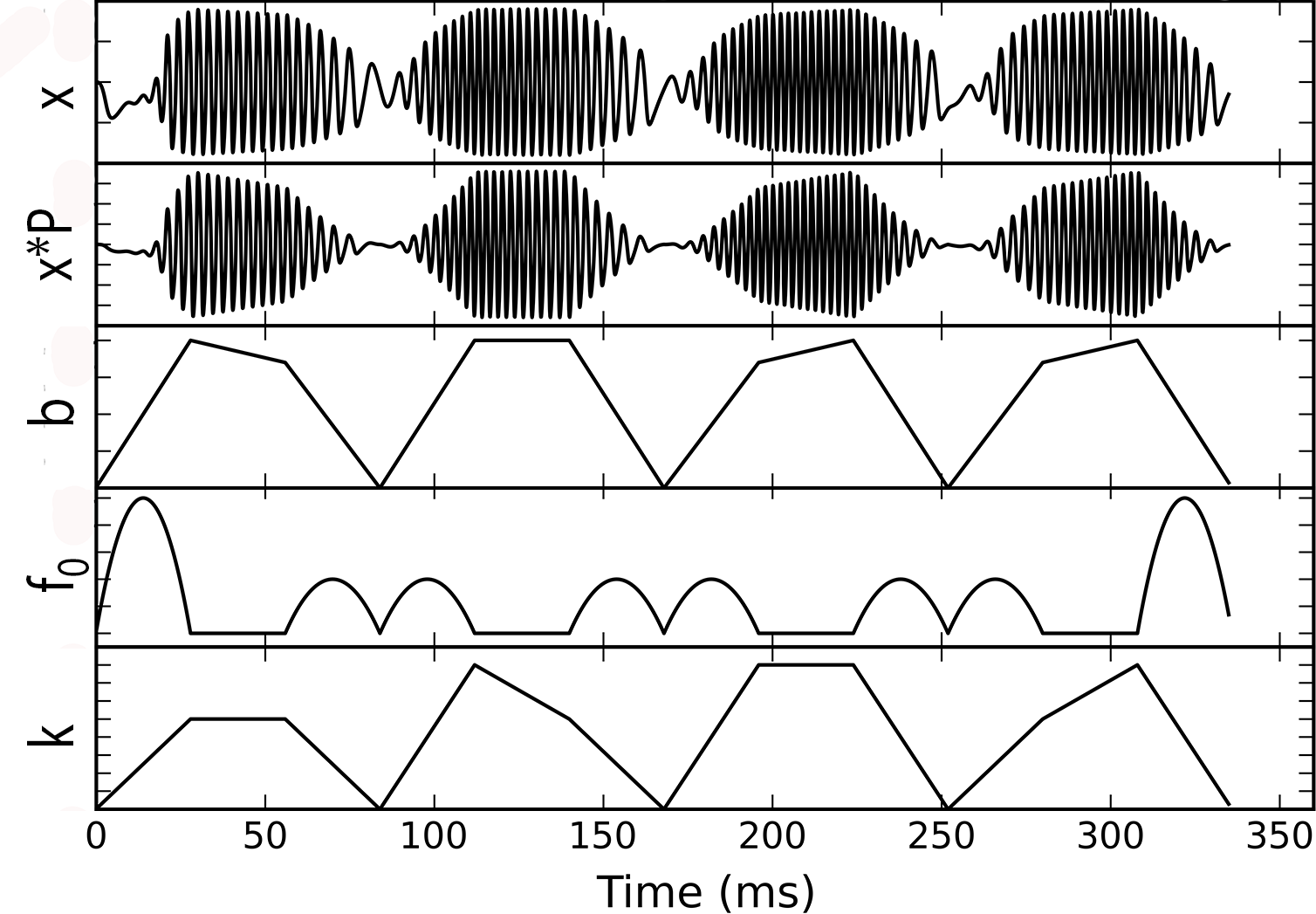} 
\includegraphics[width=0.4\textwidth]{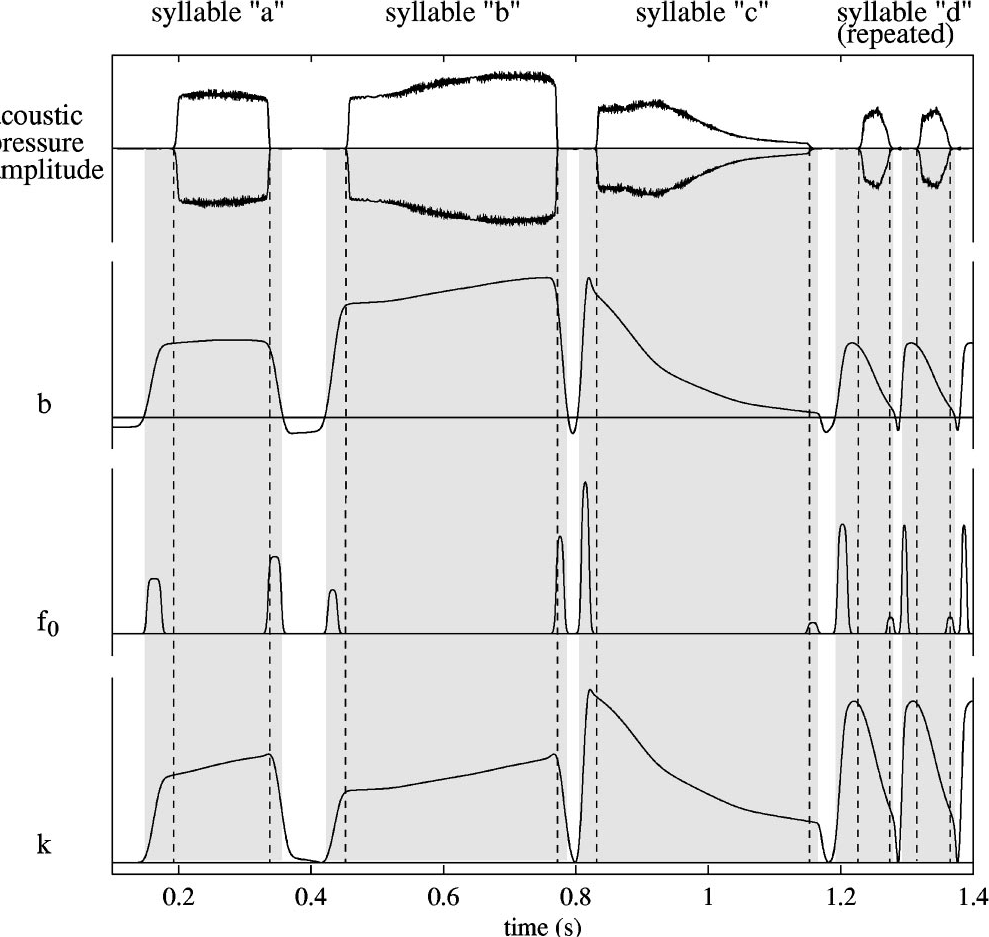} 
\caption{\textit{Left}: First panel is labial displacement x; second panel is $x(t)P(t)$, where $P$ is the air sac pressure.  Labial displacement x was driven by the time series of driving forces for song, depicted in panels 3-5: $b$, $f_0$, and $k$, respectively.  As this is an informal example, units are not provided.  \textit{Right}: Time series of driving forces reproduced from Laje et al. (2002).  These time series generated ellipses in a three-dimensional parameter space, where each ellipse represented one syllable.  The basic shape of the wave packets in the second panel at left ($(x(t)P(t)$) is intended for comparison to the top waveform at right, which shows acoustic pressure amplitude.}
\end{figure}
\begin{multicols}{2}
\noindent
ries created by Laje et al. (2002)\footnote{These were generated by creating ellipses in a three-dimensional parameter space of $k$, $b$, and $f_0$ - one for each syllable.}.  The slow modulation in the second panel at left in Figure 11 - the $x(t)P(t)$ time series - is intended to be compared to the acoustic pressure amplitude at top on the right.

Finally, note that in order to generate these driving forces, the four instructions sent by RA to the brainstem must \textit{not be independent}.  For example, to create phonation, tension $T$ and pressure $P$ must be nonzero, while $f_0$ must be effectively zero (large values of $f_0$ represent active closing of the air passageway; see Laje et al. (2002)).  The required inter-dependence may rely on cross-connectivity among the four brainstem regions, which is omitted from this model (see \textit{Discussion}).

\subsection{\textbf{Connection between driving forces and raster plot}}

Here we explain the relation between the time series of driving forces of Figure 11 (left panel) and the raster plot of Figure 8.  These time series of driving forces consist of 12 continuous segments, corresponding to the sequential firings of 12 $HVC_{RA}$ PNs.  

Let us take the leftmost RA structure of the top panel of Figure 6, which projects to rostral nXIIts (which in turn controls labial tension).  Each of the three RA PNs in this structure projects to a particular neuron in rostral nXIIts, and that neuron then encodes for a particular order regarding labial tension.  Each structure represents a triad of possible instructions.  Meanwhile, the same mechanism is at work in all four RA structures.  These four structures together can transform a signal from HVC into any of 81 possible direct commands for song.  For the complete list of rather arbitrarily-assigned instructions, and for the resulting components of song effected within each time bin, see Appendix B. 

Finally, compare the firings of the two brainstem neurons in the raster plot of Figure 8 with the time series of Figure 11, left panel.  The bursts of brainstem neuron 18, which is in rostral nXIIts, corresponds to instances on the $k(t)$ time series where $k$ achieves a particular value.  Similarly, the bursts of brainstem neuron 19, which is in PAm, corresponds to the onset of a new syllable.  It follows that the firings of some specific brainstem neurons should be found to be tightly locked - with low (\textless $\sim$ 15-ms) latency - to discontinuities in the time series of driving forces for song.

\section{DISCUSSION}
\subsection{\textbf{Model predictions}}

The model proposed in this paper makes the following predictions:
\begin{enumerate}  
  \item \textit{In RA, interneurons are less excitable than excitatory PNs, via lower resting potential}.
  \item \textit{Each RA ensemble is tightly locked to a particular $HVC_{RA}$ PN.}
  \item \textit{All $HVC_{RA}$ PNs that burst during a syllable and at gap onset extend to both rostral and ventral RA.}  There exists some evidence that certain $HVC_{RA}$ PNs target certain geographical areas in RA (Basista et al. 2014).  In particular, some HVC neurons target both dorsal and ventral RA (Kittelberger \& Mooney 1997). 
  \item \textit{Following gap onset, $HVC_{RA}$ PN activity continues throughout gaps.  While at least one $HVC_{RA}$ PN is bursting at each $\sim$ 10-ms interval, the $HVC_{RA}$ PNs bursting throughout the gap need not target vast regions of RA.}    
  \item The relative strengths of synaptic connections within HVC, within RA, HVC-to-RA, and RA-to-brainstem are roughly: 1:10:10:10, respectively.  Or: \textit{the strengths of couplings affected by neuromodulation in HVC are significantly lower than the strengths of couplings downstream}.  
  \item \textit{Specific neurons in nXIIts and RAm/PAm should be tightly locked to specific instances during song.  Those instances should be found to correspond to distinct discontinuities in the time series of a particular driving force for song} (air sac pressure $P$, labial tension $T$, or adduction force $f_0$), \textit{or to activity in PAm or Uva indicating active signaling to HVC}.  
  \item \textit{PAm projects to a region capable of effecting neurmodulatory control of HVC}; for example: VTA. 
\end{enumerate}

Regarding point 4 above: our model remains agnostic regarding the degree to which the four motor-related regions continue to be stimulated by RA throughout a gap.  Various lines of evidence, however, appear to indicate that these four regions are receiving relatively continuous information throughout gaps as well as syllables.  As noted, Leonardo \& Fee (2005) found all timescales of the song to be correlated with just one (the fastest) bursting/spiking timescale of RA ensemble activity, and that the number of RA PNs bursting per temporal location was essentially invariant.  In addition, there exists evidence that motor areas during song receive essentially continuous instructions.  Both inspiration and expiration are active processes in songbirds, even during quiet breathing (Fedde et al. 1964).  In the syrinx, even when one (lateral) side is not producing sound, the ipsilateral dorsal syringeal muscles are active to keep the syrinx closed, and ventral muscles are active on both sides (Wild 1997; Riede \& Goller 2010).  

On the other hand, Amador et al. (2013) found that the number of $HVC_{RA}$ PNs bursting during song increases at certain instances, including the onsets of syllables.  In this paper we remain agnostic regarding the activity of brainstem regions during gaps.

\subsection{\textbf{Implications for electrical stimulation studies}}

This model provides a framework within which some of the electrical stimulation studies of Ashmore et al. (2005) can be understood.  In that work, during instances following HVC stimulation when the current syllable truncated and song continued, the subsequent gap lengthened such that the total duration of song remained unchanged.  

Within the framework presented in this paper, here is what occurred.  The stimulation disrupted the ventral or rostral RA signal to nXIIts; that is: it disrupted a signal encoding a particular order for either labial tension, adduction, or both.  It did not, however, noticeably affect the signaling to respiratory regions.  

Now, in Ashmore et al. (2005), electrical stimulation only occassionally resulted in syllable truncation followed by motif continuation.  In another fraction of trials, HVC stimulations distorted but did not truncate syllables; in yet another, the motif was severed entirely.  Within the framework of this model, each of those cases represents the electrical disruption of a particular subset of the four parallel pathways from RA to brainstem - and for a particular severity of disruption.  In the case of syllable distortion, for example, pathways leading to the syrinx were affected, but less severely, than that required for syllable truncation. 

We note that the time-bin framework for HVC seems to imply that without disruption by feedback from PAm, the currently-active configuration in HVC should, upon completing a series of $HVC_{RA}$ PN firings, commence again from the beginning, replaying the first syllable - until the neurotransmitter concentration governing the inhibitory-to-inhibitory strengths $g_{ij}$ drops below some critical value.  Now, Ashmore et al. (2005) found that in a fraction of trials in which PAm stimulation truncated the motif, the motif began again from the first syllable.  This finding is consistent with the framework presented in this paper.  We do not emphasize this consistency strongly, however, given that the multiple explanations for motif restart are possible.  

\subsection{\textbf{Scaling the model}}

It will be interesting to scale the model in terms of neuron number, to examine the timing framework in finer detail.  In the small-scale model for HVC described in Figure 2, there are three $HVC_{RA}$ neurons firing per syllable-and-gap pair.  Here, the first two $HVC_{RA}$ PNs sequentially encode two notes of the syllable, and the third encodes gap onset.  That is: all three $HVC_{RA}$ neurons are used.

In a configuration with, for example, 30 $HVC_{RA}$ PNs that fire sequentially, the motif may progress as follows.  The first 20 $HVC_{RA}$ PNs encode Syllable 1 - a number that may be sufficient to permit probing the relation between $HVC_{RA}$ PN and \lq\lq note\rq\rq\ identity.  The 20th PN sends a command to terminate phonation and initiate feedback.  Until feedback reaches HVC from PAm, the PNs continue to fire, now encoding in part for the inter-syllable gap that follows Syllable 1.  Perhaps $HVC_{RA}$ PNs 21-25th are afforded the chance to fire, before feedback reaches HVC (and $HVC_{RA}$ PNs 26-30 never fire).  There exists evidence that some $HVC_{RA}$ PNs do not burst during adult song (private communication, 2016).

Further, a larger-scale model will permit the examination of an alternative to the framework presented in this paper wherein each syllable-gap pair is represented by a distinct architecture.  Specifically, we are interested in examining an HVC model that contains \textit{one} architecture that sequentially assumes distinct modes of activity - where each mode encodes the timing for a particular syllable-gap pair.

\subsection{\textbf{A concern with the \lq\lq on\rq\rq/\lq\lq off\rq\rq\ coordination of syllables}}

We identify a glaring problem with the mechanism for generating a full motif with this model.  Following an inter-syllable gap, \textit{how} does the electrical \lq\lq off\rq\rq\ signal to the currently-active configuration in HVC (Figure 3) coincide with a new chemical \lq\lq on\rq\rq\ signal to the ensuing configuration (that is: the subsequent syllable)?  Recall that in this model we attribute the former to an electrical signal from PAm via Uva, and the latter to some coordination with a region of the brain capable of modulating neurotransmitter concentrations, such as VTA - which is known to project to HVC.  For our model to work, those two signals must be essentially simultaneous - or must be separated by no more than $\sim$ 16 ms - the typical standard deviation of gap duration (Glaze \& Troyer 2006).  A direct projection from PAm to VTA, for example, would be convenient.  One has not been identified, but we note that in mammals PAm indirectly accesses brainstem regions - specifically: raphe nuclei - that are known to have neurmodulatory effects upon distant CNS regions\footnote{On a related note: we are currently examining an alternative framework in which there exists \textit{one} architecture in all of HVC, rather than a succession, which repeatedly becomes initiated into different functional configurations.}. 

\subsection{\textbf{Building cross-connectivity in RA and brainstem}}

In Figure 11 (with details in Appendix B), we showed how particular combinations of orders to the brainstem can result in time series of the driving forces for song: air sac pressure $P$, labial tension $T$ and adducting force $f_0$.  To effect song, these orders cannot be mutually independent.  Meanwhile, we took the four structures in RA that signal the brainstem to be unconnected, and those four brainstem regions to be unconnected.  It is likely that those ignored connections play a role in coordinating the driving forces.

There exists extensive evidence for dense connections at the brainstem level (e.g. Reinke \& Wild 1998, Wild 2004).  Indeed, Vicario (1991a) described the identified anatomical connectivities below RA as a \lq\lq cascade\rq\rq.  This literature has been reviewed by Schmidt \& Wild (2014), who emphasized the need to record from and manipulate distinct brainstem regions during song.  

The brainstem region is difficult to target, and for that reason extremely little is known about neuromodulatory processes and receptor dynamics downstream of RA\footnote{See Sturdy et al. (2003) regarding receptor dynamics between RA and brainstem regions, and Kubke et al. (2005) for connections between the respiratory-related brainstem regions and nXIIts.}  Largely-open questions include: 1) What are the relative contributions to song structure from RA and from the brainstem?  2) What is the role of connectivity across RA regions?  3) What is the role of connectivity between respiratory areas and nXIIts?  4) What are the relative contributions to song output from the nervous system versus motor structures?  (Regarding this last question, Mindlin (2017b) offers a review.)

Computational modeling of within-brainstem connectivity (e.g. Trevisan et al. 2006) may both complement and guide future experimental design to illuminate these questions.  A next step in our work will be to build upon the model presented in this paper various designs for cross-brainstem and cross-RA connectivity, and to assign various degrees of control to specific regions.  Of keen interest is ascertaining the minimum required model components for creating a synthetic song that HVC of an awake bird will recognize as bird's own song.

Finally, we have ignored the cross-hemisphere coordination that occurs during normal song production (Vu et al. 1998, Ashmore et al. 2008, reviewed by Schmidt \& Wild 2014).  Birds can sing with just one HVC hemisphere (Ashmore et al. 2004), however, and juveniles with one RA hemisphere destroyed at birth can develop normal song (Ashmore et al. 2008).  Lateralization at the motor level has been examined by Suthers et al. (1997), who commented that the degree of coordination is likely to be highly variable across species.  Moreover, a unilateral model is justified by the literature, although bilateral models offer the opportunity to probe more deeply the complex coordinated efforts of the CNS, in addition to identifying processes that represent redundancy.

\subsection{\textbf{The pressure-tension relation and synchronization}}

Finally, we comment on studies of pressure-tension trajectories and of synchronization in the song motor pathway, neither of which our model addresses.

Amador et al. (2013) found that both $HVC_{RA}$ PN and HVC interneuron activity was synchronized with motor instances called gestures.  A gesture (Gardner et al. 2001) is a relatively continuous trajectory in the parameter space defined by air sac pressure $P$(t) and labial tension $T$(t), and within the gesture framework song generation is described dynamically in terms of the relation between these quantities.  Further, Amador et al. (2013) found that onsets and offsets of gestures correlated - with near-zero latency - with both $HVC_{RA}$ PN and interneuron activity.  Those authors took this finding as an apparent violation of causality if one is to assume the \lq\lq clock model\rq\rq\ - in which timing in the motor pathway is set by a simple underlying clock.

Regardless of the significance of these gesture extrema: the synchronization observed by Amador et al. (2013) is not necessarily a problem for causality.  It has been shown that a dynamical system, once initiated, may rapidly converge to an attractor state.  Such a phenomenon has been modeled as \rq\rq anticipated synchronization\rq\rq\ (Matias et al. 2011; Matias et al. 2015).  

Within the context of the songbird, anticipated synchronization works as follows (Yu \& Margoliash 1996).  During the introductory notes that precede the motif, the sequence of events throughout the motor pathway indeed reflects the time delays set by causally-related regions.  By the time of onset of the actual motif, however, these regions have become synchronized.  That is: a connectivity such as that described in this paper is required to initiate song, but it may be dispensed with for some duration thereafter, once the attractor state is reached.  

In future modeling, we will examine: 1) implications regarding relationships among $P$ and $T$ and other possibly-significant dynamical quantities; 2) the synaptic connections that may be required to incite synchronization in the song motor pathway.

\section{LOOKING FORWARD}
In closing, we pose questions that may be probed via a combination of computation modeling and the design of new experiments.  1) At what geographical location(s) in the song motor pathway does \lq\lq note\rq\rq\ acquire definition?  2) Can a note be further divided?  3) How do cross-brainstem and inter-hemisphere connectivity affect acoustic output?  4) Which CNS regions must be causally connected (via electrical synapses) in order to incite the synchronization of distinct nuclei during song?  5) How biophysically detailed must a neuronal network model be in order to produce synthetic songs that an animal of the species will recognize?  We look forward to expanding the simple model set forth in this paper, to determine whether it proves useful for probing these fascinating problems.

\section{ACKNOWLEDGEMENTS}
Thank you to Marc Schmidt for invaluable guidance in shaping this paper.  Thanks also to Franz Goller, Anthony Leonardo, Daniel Margoliash, Gabriel Mindlin, Ofer Tchernichovski, and Martin Wild for informative conversations.

\end{multicols}
\bibliographystyle{acm}
\nocite{*}
\bibliography{refs}

\begin{thebibliography}{10}

\bibitem{abarbanel2004mapping}
{\sc Abarbanel, H.~D., Gibb, L., Mindlin, G.~B., and Talathi, S.}
\newblock Mapping neural architectures onto acoustic features of birdsong.
\newblock {\em Journal of neurophysiology 92}, 1 (2004), 96--110.

\bibitem{albert1996temporal}
{\sc Albert, C.~Y., and Margoliash, D.}
\newblock Temporal hierarchical control of singing in birds.
\newblock {\em Science 273}, 5283 (1996), 1871.

\bibitem{alonso2016integrated}
{\sc Alonso, R.~G., Amador, A., and Mindlin, G.~B.}
\newblock An integrated model for motor control of song in serinus canaria.
\newblock {\em Journal of Physiology-Paris\/} (2016).

\bibitem{alonso2015circular}
{\sc Alonso, R.~G., Trevisan, M.~A., Amador, A., Goller, F., and Mindlin,
  G.~B.}
\newblock A circular model for song motor control in serinus canaria.
\newblock {\em Frontiers in computational neuroscience 9\/} (2015).

\bibitem{amador2008beyond}
{\sc Amador, A., and Mindlin, G.~B.}
\newblock Beyond harmonic sounds in a simple model for birdsong production.
\newblock {\em Chaos: An Interdisciplinary Journal of Nonlinear Science 18}, 4
  (2008), 043123.

\bibitem{amador2013elemental}
{\sc Amador, A., Perl, Y.~S., Mindlin, G., and Margoliash, D.}
\newblock Elemental gesture dynamics are encoded by song premotor cortical
  neurons.
\newblock {\em Nature 495}, 7439 (2013), 59.

\bibitem{andalman2011control}
{\sc Andalman, A.~S., Foerster, J.~N., and Fee, M.~S.}
\newblock Control of vocal and respiratory patterns in birdsong: dissection of
  forebrain and brainstem mechanisms using temperature.
\newblock {\em PLoS One 6}, 9 (2011), e25461.

\bibitem{armstrong2016model}
{\sc Armstrong, E., and Abarbanel, H.~D.}
\newblock Model of the songbird nucleus hvc as a network of central pattern
  generators.
\newblock {\em Journal of neurophysiology 116}, 5 (2016), 2405--2419.

\bibitem{aronov2008specialized}
{\sc Aronov, D., Andalman, A.~S., and Fee, M.~S.}
\newblock A specialized forebrain circuit for vocal babbling in the juvenile
  songbird.
\newblock {\em Science 320}, 5876 (2008), 630--634.

\bibitem{ashmore2008hemispheric}
{\sc Ashmore, R.~C., Bourjaily, M., and Schmidt, M.~F.}
\newblock Hemispheric coordination is necessary for song production in adult
  birds: implications for a dual role for forebrain nuclei in vocal motor
  control.
\newblock {\em Journal of neurophysiology 99}, 1 (2008), 373--385.

\bibitem{ashmore2005brainstem}
{\sc Ashmore, R.~C., Wild, J.~M., and Schmidt, M.~F.}
\newblock Brainstem and forebrain contributions to the generation of learned
  motor behaviors for song.
\newblock {\em Journal of Neuroscience 25}, 37 (2005), 8543--8554.

\bibitem{basista2014independent}
{\sc Basista, M.~J., Elliott, K.~C., Wu, W., Hyson, R.~L., Bertram, R., and
  Johnson, F.}
\newblock Independent premotor encoding of the sequence and structure of
  birdsong in avian cortex.
\newblock {\em Journal of Neuroscience 34}, 50 (2014), 16821--16834.

\bibitem{beckers2003pure}
{\sc Beckers, G.~J., Suthers, R.~A., and Ten~Cate, C.}
\newblock Pure-tone birdsong by resonance filtering of harmonic overtones.
\newblock {\em Proceedings of the National Academy of Sciences 100}, 12 (2003),
  7372--7376.

\bibitem{bottjer1984forebrain}
{\sc Bottjer, S., Miesner, E.~A., and Arnold, A.~P.}
\newblock Forebrain lesions disrupt development but not maintenance of song in
  passerine birds.
\newblock {\em Science 224\/} (1984), 901--904.

\bibitem{brainard2002songbirds}
{\sc Brainard, M.~S., and Doupe, A.~J.}
\newblock What songbirds teach us about learning.
\newblock {\em Nature 417}, 6886 (2002), 351--358.

\bibitem{cannon2015neural}
{\sc Cannon, J., Kopell, N., Gardner, T., and Markowitz, J.}
\newblock Neural sequence generation using spatiotemporal patterns of
  inhibition.
\newblock {\em PLoS Comput Biol 11}, 11 (2015), e1004581.

\bibitem{daou2013electrophysiological}
{\sc Daou, A., Ross, M.~T., Johnson, F., Hyson, R.~L., and Bertram, R.}
\newblock Electrophysiological characterization and computational models of hvc
  neurons in the zebra finch.
\newblock {\em Journal of neurophysiology 110}, 5 (2013), 1227--1245.

\bibitem{destexhe1994synthesis}
{\sc Destexhe, A., Mainen, Z.~F., and Sejnowski, T.~J.}
\newblock Synthesis of models for excitable membranes, synaptic transmission
  and neuromodulation using a common kinetic formalism.
\newblock {\em Journal of computational neuroscience 1}, 3 (1994), 195--230.

\bibitem{destexhe2001thalamocortical}
{\sc Destexhe, A., and Sejnowski, T.~J.}
\newblock Thalamocortical assemblies: How ion channels, single neurons and
  large-scale networks organize sleep oscillations.

\bibitem{elemans2008superfast}
{\sc Elemans, C.~P., Mead, A.~F., Rome, L.~C., and Goller, F.}
\newblock Superfast vocal muscles control song production in songbirds.
\newblock {\em PloS one 3}, 7 (2008), e2581.

\bibitem{fedde1964electromyographic}
{\sc Fedde, M., Burger, R.~E., and Kitchell, R.}
\newblock Electromyographic studies of the effects of bodily position and
  anesthesia on the activity of the respiratory muscles of the domestic cock.
\newblock {\em Poultry Science 43}, 4 (1964), 839--846.

\bibitem{fee2004neural}
{\sc Fee, M.~S., Kozhevnikov, A.~A., and Hahnloser, R.~H.}
\newblock Neural mechanisms of vocal sequence generation in the songbird.
\newblock {\em Annals of the New York Academy of Sciences 1016}, 1 (2004),
  153--170.

\bibitem{fee1998role}
{\sc Fee, M.~S., Shraiman, B., Pesaran, B., and Mitra, P.~P.}
\newblock The role of nonlinear dynamics of the syrinx in the vocalizations of
  a songbird.
\newblock {\em Nature 395}, 6697 (1998), 67.

\bibitem{gale2006physiological}
{\sc Gale, S.~D., and Perkel, D.~J.}
\newblock Physiological properties of zebra finch ventral tegmental area and
  substantia nigra pars compacta neurons.
\newblock {\em Journal of neurophysiology 96}, 5 (2006), 2295--2306.

\bibitem{gardner2001simple}
{\sc Gardner, T., Cecchi, G., Magnasco, M., Laje, R., and Mindlin, G.~B.}
\newblock Simple motor gestures for birdsongs.
\newblock {\em Physical review letters 87}, 20 (2001), 208101.

\bibitem{gibb2009inhibition}
{\sc Gibb, L., Gentner, T.~Q., and Abarbanel, H.~D.}
\newblock Inhibition and recurrent excitation in a computational model of
  sparse bursting in song nucleus hvc.
\newblock {\em Journal of neurophysiology 102}, 3 (2009a), 1748--1762.

\bibitem{gibb2009brain}
{\sc Gibb, L., Gentner, T.~Q., and Abarbanel, H.~D.}
\newblock Brain stem feedback in a computational model of birdsong sequencing.
\newblock {\em Journal of neurophysiology 102}, 3 (2009b), 1763--1778.

\bibitem{glaze2006temporal}
{\sc Glaze, C.~M., and Troyer, T.~W.}
\newblock Temporal structure in zebra finch song: implications for motor
  coding.
\newblock {\em Journal of Neuroscience 26}, 3 (2006), 991--1005.

\bibitem{goller1996role}
{\sc Goller, F., and Suthers, R.~A.}
\newblock Role of syringeal muscles in gating airflow and sound production in
  singing brown thrashers.
\newblock {\em Journal of Neurophysiology 75}, 2 (1996), 867--876.

\bibitem{hahnloser2002ultra}
{\sc Hahnloser, R.~H., Kozhevnikov, A.~A., and Fee, M.~S.}
\newblock An ultra-sparse code underlies the generation of neural sequences in
  a songbird.
\newblock {\em Nature 419}, 6902 (2002), 65.

\bibitem{hamaguchi2012recurrent}
{\sc Hamaguchi, K., and Mooney, R.}
\newblock Recurrent interactions between the input and output of a songbird
  cortico-basal ganglia pathway are implicated in vocal sequence variability.
\newblock {\em Journal of Neuroscience 32}, 34 (2012), 11671--11687.

\bibitem{izhikevich2007dynamical}
{\sc Izhikevich, E.~M.}
\newblock {\em Dynamical systems in neuroscience}.
\newblock MIT press, 2007.

\bibitem{jensen2007songbirds}
{\sc Jensen, K.~K., Cooper, B.~G., Larsen, O.~N., and Goller, F.}
\newblock Songbirds use pulse tone register in two voices to generate
  low-frequency sound.
\newblock {\em Proceedings of the Royal Society of London B: Biological
  Sciences 274}, 1626 (2007), 2703--2710.

\bibitem{kittelberger1997individual}
{\sc Kittelberger, M., and Mooney, R.}
\newblock Individual hvc axons innervate ra subdomains that control temporal
  and spectral elements of learned song.
\newblock In {\em Soc Neurosci Abstr\/} (1997), vol.~23.

\bibitem{kosche2015interplay}
{\sc Kosche, G., Vallentin, D., and Long, M.~A.}
\newblock Interplay of inhibition and excitation shapes a premotor neural
  sequence.
\newblock {\em Journal of Neuroscience 35}, 3 (2015), 1217--1227.

\bibitem{kozhevnikov2007singing}
{\sc Kozhevnikov, A.~A., and Fee, M.~S.}
\newblock Singing-related activity of identified hvc neurons in the zebra
  finch.
\newblock {\em Journal of neurophysiology 97}, 6 (2007), 4271--4283.

\bibitem{kubota1991nmda}
{\sc Kubota, M., and Saito, N.}
\newblock Nmda receptors participate differentially in two different synaptic
  inputs in neurons of the zebra finch robust nucleus of the archistriatum in
  vitro.
\newblock {\em Neuroscience letters 125}, 2 (1991), 107--109.

\bibitem{laje2002neuromuscular}
{\sc Laje, R., Gardner, T.~J., and Mindlin, G.~B.}
\newblock Neuromuscular control of vocalizations in birdsong: a model.
\newblock {\em Physical Review E 65}, 5 (2002), 051921.

\bibitem{larsen2002direct}
{\sc Larsen, O.~N., and Goller, F.}
\newblock Direct observation of syringeal muscle function in songbirds and a
  parrot.
\newblock {\em Journal of Experimental Biology 205}, 1 (2002), 25--35.

\bibitem{leonardo2005ensemble}
{\sc Leonardo, A., and Fee, M.~S.}
\newblock Ensemble coding of vocal control in birdsong.
\newblock {\em Journal of Neuroscience 25}, 3 (2005), 652--661.

\bibitem{lewandowski2013interface}
{\sc Lewandowski, B., Vyssotski, A., Hahnloser, R.~H., and Schmidt, M.}
\newblock At the interface of the auditory and vocal motor systems: Nif and its
  role in vocal processing, production and learning.
\newblock {\em Journal of Physiology-Paris 107}, 3 (2013), 178--192.

\bibitem{li2006stable}
{\sc Li, M., and Greenside, H.}
\newblock Stable propagation of a burst through a one-dimensional homogeneous
  excitatory chain model of songbird nucleus hvc.
\newblock {\em Physical Review E 74}, 1 (2006), 011918.

\bibitem{long2010support}
{\sc Long, M.~A., Jin, D.~Z., and Fee, M.~S.}
\newblock Support for a synaptic chain model of neuronal sequence generation.
\newblock {\em Nature 468}, 7322 (2010), 394--399.

\bibitem{lynch2016rhythmic}
{\sc Lynch, G.~F., Okubo, T.~S., Hanuschkin, A., Hahnloser, R.~H., and Fee,
  M.~S.}
\newblock Rhythmic continuous-time coding in the songbird analog of vocal motor
  cortex.
\newblock {\em Neuron 90}, 4 (2016), 877--892.

\bibitem{margoliash1997functional}
{\sc Margoliash, D.}
\newblock Functional organization of forebrain pathways for song production and
  perception.
\newblock {\em Developmental Neurobiology 33}, 5 (1997), 671--693.

\bibitem{matias2011anticipated}
{\sc Matias, F.~S., Carelli, P.~V., Mirasso, C.~R., and Copelli, M.}
\newblock Anticipated synchronization in a biologically plausible model of
  neuronal motifs.
\newblock {\em Physical Review E 84}, 2 (2011), 021922.

\bibitem{matias2015self}
{\sc Matias, F.~S., Carelli, P.~V., Mirasso, C.~R., and Copelli, M.}
\newblock Self-organized near-zero-lag synchronization induced by spike-timing
  dependent plasticity in cortical populations.
\newblock {\em PloS one 10}, 10 (2015), e0140504.

\bibitem{mclean2013characterization}
{\sc McLean, J., Bricault, S., and Schmidt, M.~F.}
\newblock Characterization of respiratory neurons in the rostral ventrolateral
  medulla, an area critical for vocal production in songbirds.
\newblock {\em Journal of neurophysiology 109}, 4 (2013), 948--957.

\bibitem{mindlin2017nonlinear}
{\sc Mindlin, G.~B.}
\newblock Nonlinear dynamics in the study of birdsong.
\newblock {\em Chaos: An Interdisciplinary Journal of Nonlinear Science 27}, 9
  (2017a), 092101.

\bibitem{mindlin2017avian}
{\sc Mindlin, G.~B.}
\newblock Avian vocal production beyond low dimensional models.
\newblock {\em Journal of Statistical Mechanics: Theory and Experiment 2017}, 2
  (2017b), 024005.

\bibitem{mooney2009neurobiology}
{\sc Mooney, R.}
\newblock Neurobiology of song learning.
\newblock {\em Current opinion in neurobiology 19}, 6 (2009), 654--660.

\bibitem{mooney1991two}
{\sc Mooney, R., and Konishi, M.}
\newblock Two distinct inputs to an avian song nucleus activate different
  glutamate receptor subtypes on individual neurons.
\newblock {\em Proceedings of the National Academy of Sciences 88}, 10 (1991),
  4075--4079.

\bibitem{nottebohm1976central}
{\sc Nottebohm, F., Stokes, T.~M., and Leonard, C.~M.}
\newblock Central control of song in the canary, serinus canarius.
\newblock {\em Journal of Comparative Neurology 165}, 4 (1976), 457--486.

\bibitem{perl2011reconstruction}
{\sc Perl, Y.~S., Arneodo, E.~M., Amador, A., Goller, F., and Mindlin, G.~B.}
\newblock Reconstruction of physiological instructions from zebra finch song.
\newblock {\em Physical Review E 84}, 5 (2011), 051909.

\bibitem{picardo2016population}
{\sc Picardo, M.~A., Merel, J., Katlowitz, K.~A., Vallentin, D., Okobi, D.~E.,
  Benezra, S.~E., Clary, R.~C., Pnevmatikakis, E.~A., Paninski, L., and Long,
  M.~A.}
\newblock Population-level representation of a temporal sequence underlying
  song production in the zebra finch.
\newblock {\em Neuron 90}, 4 (2016), 866--876.

\bibitem{reinke1998identification}
{\sc Reinke, H., and Wild, J.}
\newblock Identification and connections of inspiratory premotor neurons in
  songbirds and budgerigar.
\newblock {\em Journal of Comparative Neurology 391}, 2 (1998), 147--163.

\bibitem{riede2010peripheral}
{\sc Riede, T., and Goller, F.}
\newblock Peripheral mechanisms for vocal production in birds--differences and
  similarities to human speech and singing.
\newblock {\em Brain and language 115}, 1 (2010), 69--80.

\bibitem{roberts2008telencephalic}
{\sc Roberts, T.~F., Klein, M.~E., Kubke, M.~F., Wild, J.~M., and Mooney, R.}
\newblock Telencephalic neurons monosynaptically link brainstem and forebrain
  premotor networks necessary for song.
\newblock {\em Journal of Neuroscience 28}, 13 (2008), 3479--3489.

\bibitem{schmidt2004bilateral}
{\sc Schmidt, M.~F., Ashmore, R.~C., and Vu, E.~T.}
\newblock Bilateral control and interhemispheric coordination in the avian song
  motor system.
\newblock {\em Annals of the New York Academy of Sciences 1016}, 1 (2004),
  171--186.

\bibitem{schmidt2014respiratory}
{\sc Schmidt, M.~F., and Wild, J.~M.}
\newblock The respiratory-vocal system of songbirds: anatomy, physiology, and
  neural control.
\newblock {\em Progress in brain research 212\/} (2014), 297.

\bibitem{simpson1990brain}
{\sc Simpson, H.~B., and Vicario, D.~S.}
\newblock Brain pathways for learned and unlearned vocalizations differ in
  zebra finches.
\newblock {\em Journal of Neuroscience 10}, 5 (1990), 1541--1556.

\bibitem{sitt2008dynamical}
{\sc Sitt, J., Amador, A., Goller, F., and Mindlin, G.}
\newblock Dynamical origin of spectrally rich vocalizations in birdsong.
\newblock {\em Physical Review E 78}, 1 (2008), 011905.

\bibitem{spiro1999long}
{\sc Spiro, J.~E., Dalva, M.~B., and Mooney, R.}
\newblock Long-range inhibition within the zebra finch song nucleus ra can
  coordinate the firing of multiple projection neurons.
\newblock {\em Journal of neurophysiology 81}, 6 (1999), 3007--3020.

\bibitem{striedter1998bilateral}
{\sc Striedter, G., and Vu, E.}
\newblock Bilateral feedback projections to the forebrain in the premotor
  network for singing in zebra finches.
\newblock {\em Developmental Neurobiology 34}, 1 (1998), 27--40.

\bibitem{suthers1997peripheral}
{\sc Suthers, R.~A.}
\newblock Peripheral control and lateralization of birdsong.
\newblock {\em Developmental Neurobiology 33}, 5 (1997), 632--652.

\bibitem{trevisan2006nonlinear}
{\sc Trevisan, M.~A., Mindlin, G.~B., and Goller, F.}
\newblock Nonlinear model predicts diverse respiratory patterns of birdsong.
\newblock {\em Physical review letters 96}, 5 (2006), 058103.

\bibitem{vallentin2016inhibition}
{\sc Vallentin, D., Kosche, G., Lipkind, D., and Long, M.~A.}
\newblock Inhibition protects acquired song segments during vocal learning in
  zebra finches.
\newblock {\em Science 351}, 6270 (2016), 267--271.

\bibitem{vates1996auditory}
{\sc Vates, G.~E., Broome, B.~M., Mello, C.~V., and Nottebohm, F.}
\newblock Auditory pathways of caudal telencephalon and their relation to the
  song system of adult male zebra finches (taenopygia guttata).
\newblock {\em Journal of Comparative Neurology 366}, 4 (1996), 613--642.

\bibitem{verduzco2012model}
{\sc Verduzco-Flores, S.~O., Bodner, M., and Ermentrout, B.}
\newblock A model for complex sequence learning and reproduction in neural
  populations.
\newblock {\em Journal of computational neuroscience 32}, 3 (2012), 403--423.

\bibitem{vicario1991organization}
{\sc Vicario, D.~S.}
\newblock Organization of the zebra finch song control system: functional
  organization of outputs from nucleus robustus archistriatalis.
\newblock {\em Journal of Comparative Neurology 309}, 4 (1991a), 486--494.

\bibitem{vicario1991contributions}
{\sc Vicario, D.~S.}
\newblock Contributions of syringeal muscles to respiration and vocalization in
  the zebra finch.
\newblock {\em Developmental Neurobiology 22}, 1 (1991b), 63--73.

\bibitem{vu1994identification}
{\sc Vu, E.~T., Mazurek, M.~E., and Kuo, Y.-C.}
\newblock Identification of a forebrain motor programming network for the
  learned song of zebra finches.
\newblock {\em Journal of Neuroscience 14}, 11 (1994), 6924--6934.

\bibitem{wild2004pulmonary}
{\sc Wild, J.}
\newblock Pulmonary and tracheosyringeal afferent inputs to the avian song
  system.
\newblock In {\em Seventh Congress of the International Society for
  Neuroethology, Nyborg, Denmark\/} (2004), p.~070.

\bibitem{wild1998inspiratory}
{\sc Wild, J., Goller, F., and Suthers, R.}
\newblock Inspiratory muscle activity during bird song.
\newblock {\em Journal of neurobiology 36}, 3 (1998), 441.

\bibitem{wild1997projections}
{\sc Wild, J., Li, D., and Eagleton, C.}
\newblock Projections of the dorsomedial nucleus of the intercollicular complex
  (dm) in relation to respiratory-vocal nuclei in the brainstem of pigeon
  (columba livia) and zebra finch (taeniopygia guttata).
\newblock {\em Journal of Comparative Neurology 377}, 3 (1997), 392--413.

\bibitem{wild1993descending}
{\sc Wild, J.~M.}
\newblock Descending projections of the songbird nucleus robustus
  archistriatalis.
\newblock {\em Journal of Comparative Neurology 338}, 2 (1993a), 225--241.

\bibitem{wild1993avian}
{\sc Wild, J.~M.}
\newblock The avian nucleus retroambigualis: a nucleus for breathing, singing
  and calling.
\newblock {\em Brain research 606}, 2 (1993b), 319--324.

\bibitem{wild1997neural}
{\sc Wild, J.~M.}
\newblock Neural pathways for the control of birdsong production.
\newblock {\em Developmental Neurobiology 33}, 5 (1997), 653--670.

\bibitem{yildiz2011hierarchical}
{\sc Yildiz, I.~B., and Kiebel, S.~J.}
\newblock A hierarchical neuronal model for generation and online recognition
  of birdsongs.
\newblock {\em PLoS Computational Biology 7}, 12 (2011), e1002303.

\end{thebibliography}

\begin{multicols}{2}
\section{Appendix A: Mechanism for HVC-RA and RA-to-brainstem interactions}
Here we present details regarding the mechanism by which an $HVC_{RA}$ PN recruits an ensemble of RA PNs, and how those RA PNs in turn recruit brainstem neurons.  This procedure was used to generate the raster plot of Figure 8.  

\subsection{\textbf{RA electrophysiology and connectivity reproduce observed RA activity during quiescence.}}

As described in \textit{Model}, RA contains four identical structures, represented in Figure 12.  We first mimic quiescence in RA immediately prior to and following a song.  Here the excitatory and inhibitory populations receive a low background excitation but no direct excitation from HVC, and the interneurons are rendered less excitable than the RA PNs, via the reversal potential of leak current ($E_{L,i}$ and $E_{L,e}$ are -85 and -80 mV, respectively).  

Figure 13 shows the resulting voltage traces for one of the four structures in RA.  Numberings correspond to the numberings on the schematic of Figure 12: three inhibitory neurons (Cells  0, 1, and 2) at left, and three excitatory PNs (3, 4, and 5) at right.  The former are silent above threshold, while the latter spike or burst continually.

\subsection{\textbf{HVC RA PNs recruit RA PN ensembles by exciting RA interneurons.}}

Now we demonstrate how an excitatory signal from an $HVC_{RA}$ neuron directly excites an RA interneuron, and how that interneuron in turn suppresses a fraction of the RA PNs\footnote{Here we hold the inhibitory-to-inhibitory couplings $g_{ij}$ in HVC at constant elevated values, in order to demonstrate the robustness of the resulting relationship between HVC and RA activity.  That is: here we are not invoking the $g_{ij}$-$T_{max}$ relation that was used to construct the raster plot.}.  

Let us take as an example the top panel of the HVC-to-RA connectivity diagram of Figure 6, which corresponds to Syllable $a$ and the subsequent gap.  Within that panel, let us take the interaction between the $HVC_{RA}$ FHU and the leftmost RA structure.  Note that $HVC_{RA}$ neurons 3 and 4 directly excite RA interneuron 0.  Figure 14 shows the voltage traces of the six RA neurons shown previously in quiescence, now when connectivity from HVC is turned on.  RA interneuron 0 is now bursting.  

The top panel of Figure 15 elucidates this mechanism.  Here, the colorings on the voltage traces (left) correspond to the colorings of the schematics at right.  The burst of either HVC 3 (blue) or HVC 4 (cyan) incites a burst from RA 0 (magenta).  

The middle panel of Figure 15 shows the effect of a burst from RA interneuron 0 upon the RA PNs within that structure.  Given the inhibitory projections of RA interneuron 0 to RA PNs 3 and 5 - but not to RA PN 4: while RA interneuron 0 bursts, the bursting of RA PN 3 and RA PN 5 is suppressed - and the bursting of RA PN 4 is permitted.  Further, recall that each $HVC_{RA}$ PN has such a relationship with an interneuron in four such structures identical to that depicted in Figure 12.  It is in this way that each HVC PN - via interneuron connectivity - selects an ensemble of RA PNs.
\begin{figure}[H]
  \centering
\includegraphics[width=0.4\textwidth]{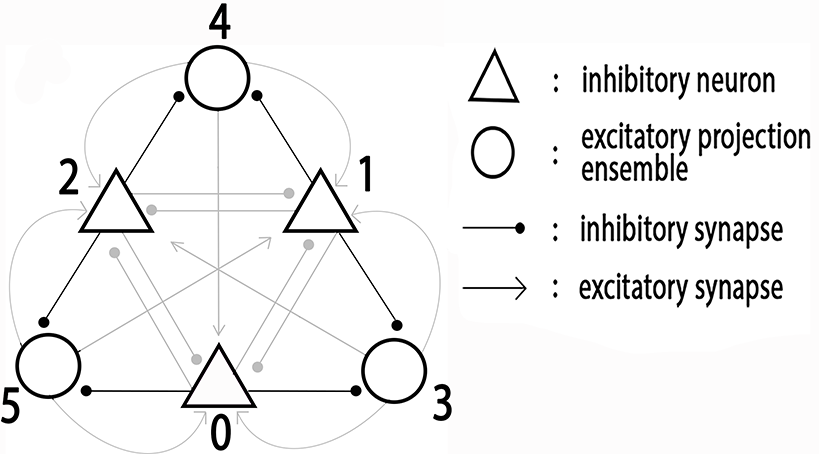} 
\caption{One of four identical six-neuron structures in RA.  Numbering corresponds to numbers on voltage traces in Figure 13.}
\end{figure}
\end{multicols}
\begin{figure}[H]
  \centering
\includegraphics[width=0.4\textwidth]{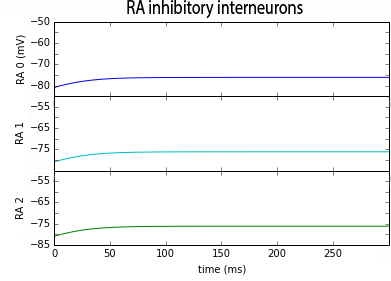}
\includegraphics[width=0.4\textwidth]{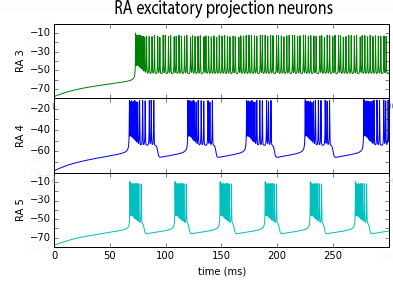} 
\caption{Activity of RA interneurons and excitatory projection neurons during quiescence, when the $HVC_{RA}$ PN population is inactive.  The RA interneurons possess a lower leak reversal potential than do the RA PNs: -85 vs -80 mV, respectively.  Consequently, when all six neurons receive the same background excitation, the interneurons are silent and the RA PNs are active.  Numbering corresponds to numbers on schematic of Figure 12.}
\end{figure}
\begin{figure}[H]
  \centering
\includegraphics[width=0.4\textwidth]{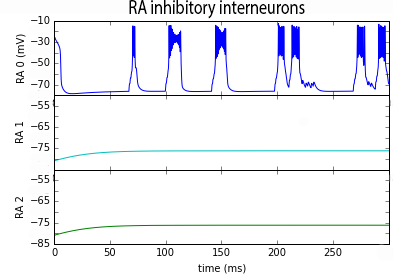} 
\includegraphics[width=0.4\textwidth]{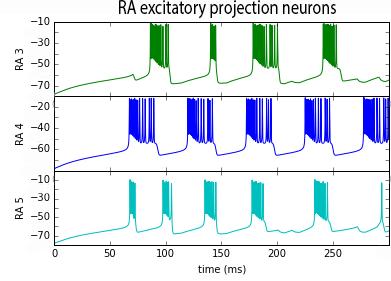} 
\caption{Activity of RA interneurons and excitatory projection neurons when $HVC_{RA}$ PNs 3 and 4 directly excite RA interneuron 0.  RA interneuron 0 now bursts.}
\end{figure}
\begin{multicols}{2}
\subsection{\textbf{The four distinct RA structures send four $\sim$ simultaneous signals to four distinct brainstem regions.}}

In the final stage of the computational model, a brainstem neuron is excited.  The bottom panel of Figure 15 is a summary.  Here, $HVC_{RA}$ neurons 3 and 4 (blue and cyan) excite RA interneuron 0 (magenta), which suppresses RA PNs 3 and 5 (green and black), but permits RA PN 4 (yellow) to burst.  RA PN 4 then excites a neuron in the brainstem (black).

\section{Appendix B: Creating the example brainstem-to-motor connectivity and driving forces for song}
\subsection{\textbf{Brainstem assignments for creating the driving forces for song}}

As noted in \textit{Results}, we assigned each of the three RA PNs in each of the four RA structures a particular order for its associated brainstem neuron.  These assignments are as follows.

For the three RA PNs in the RA structure that sends instructions to rostral nXIIts (leftmost in Figure 6), respectively: 1) \lq\lq Effect Tension Value 1\rq\rq; 2) \lq\lq Effect Tension Value 2\rq\rq; 3) \lq\lq Set labia at resting locations (Tension $=$ 0)\rq\rq.  

For the three RA PNs in the RA structure projecting to caudal nXIIts (second from left in Figure 6): 1) \lq\lq Effect
\end{multicols}
\begin{figure}[H]
  \centering
\includegraphics[width=0.5\textwidth,valign=t]{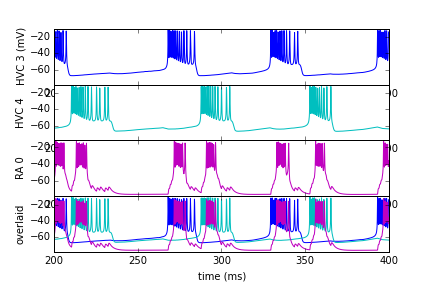}
\includegraphics[width=0.15\textwidth,valign=t]{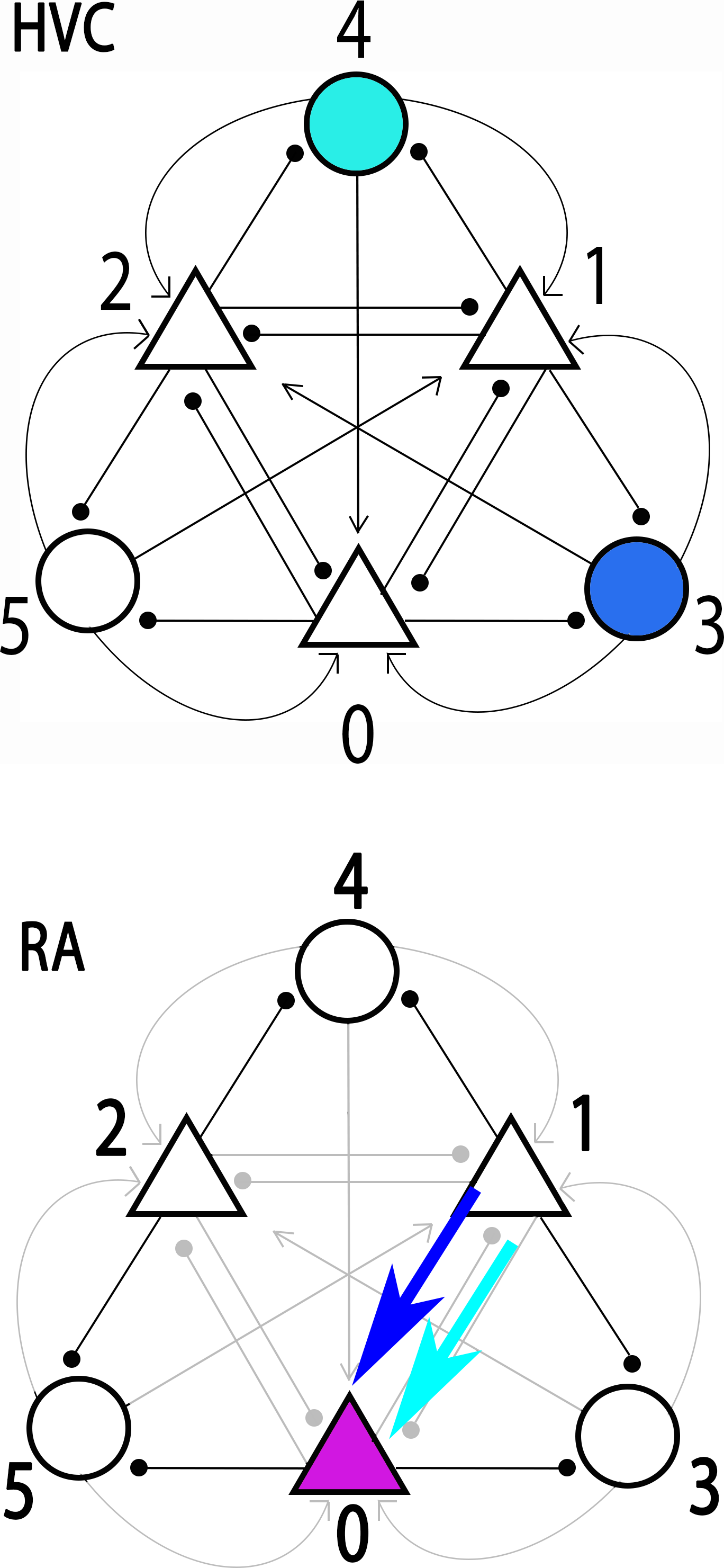}\\
\hrule
\includegraphics[width=0.5\textwidth,valign=t]{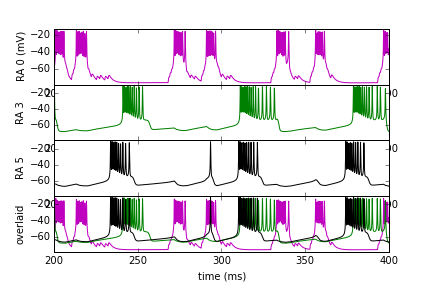}
\includegraphics[width=0.15\textwidth,valign=t]{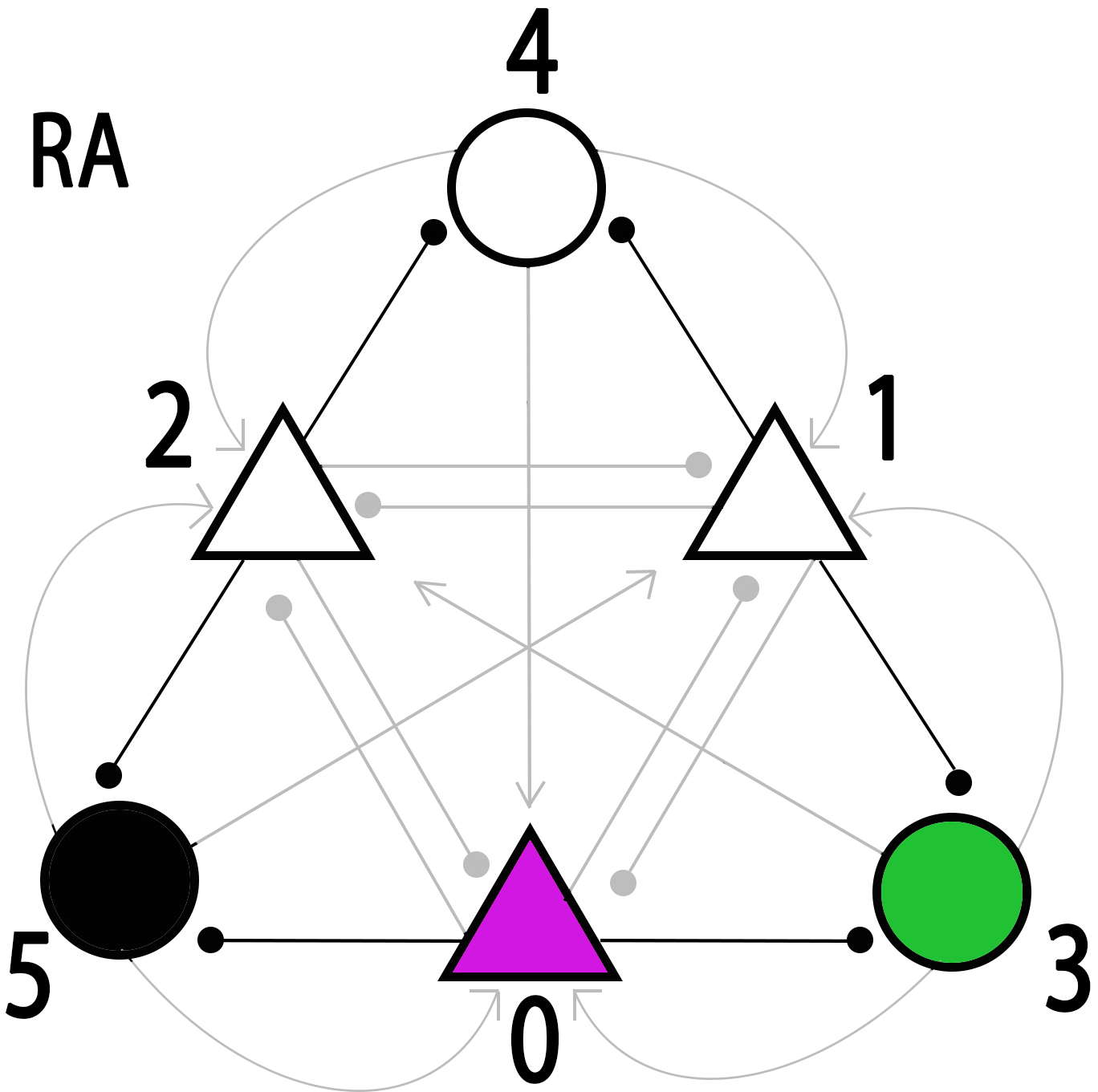} \\
\hrule
\includegraphics[width=0.5\textwidth,valign=t]{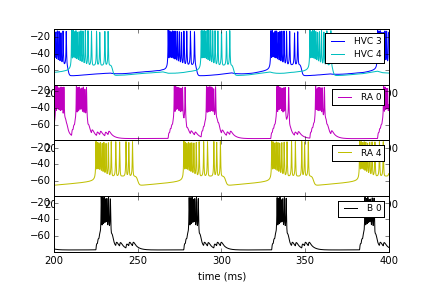}
\includegraphics[width=0.15\textwidth,valign=t]{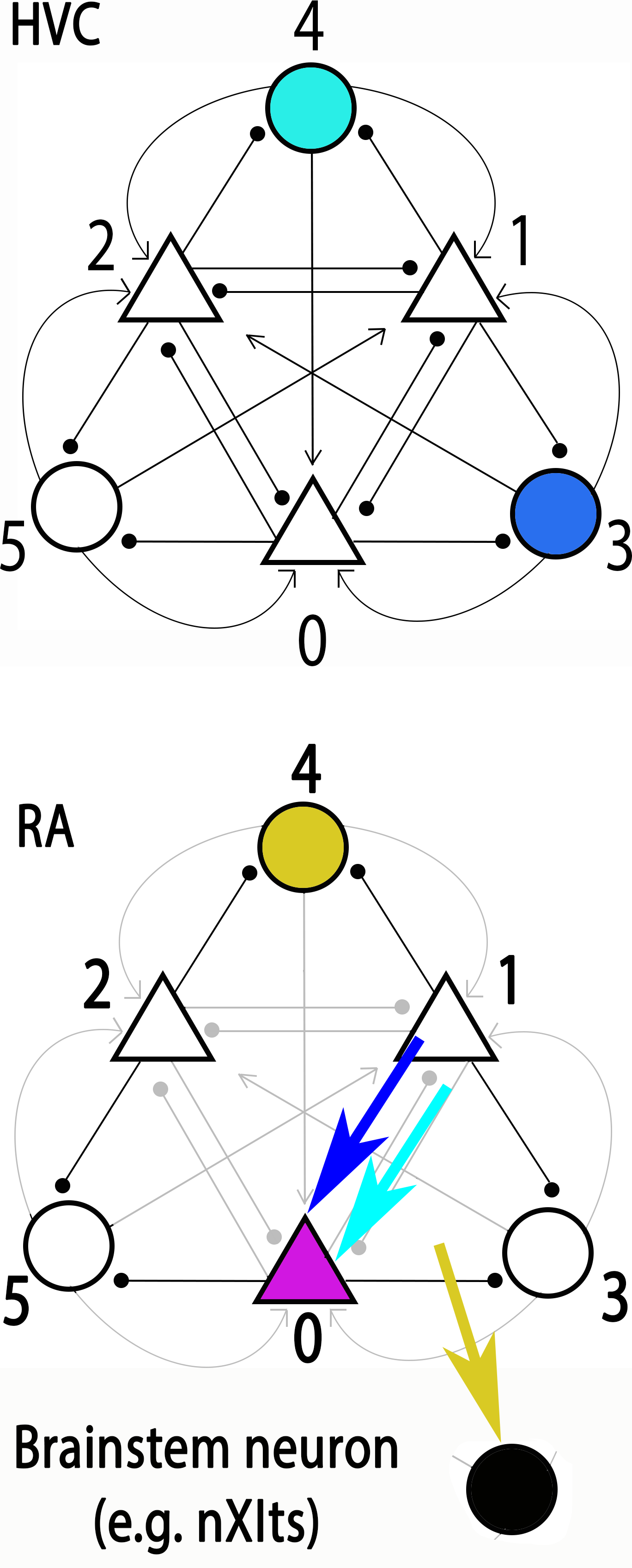}
\caption{\textbf{Three events}.  \textit{Top}: $HVC_{RA}$ PNs 3 and 4 directly excite RA interneuron 0.  \textit{Middle}: RA interneuron 0 then suppresses RA PNs 3 and 5, but not RA PN 4.  \textit{Bottom}: Summary.  $HVC_{RA}$ PNs 3 and 4 excite RA interneuron 0, which permits only RA PN 4 (and not 3 or 5) to fire.  Finally, RA PN 4 directly excites a neuron in the brainstem.}
\end{figure}
\begin{figure}[H]
  \centering
\includegraphics[width=0.9\textwidth]{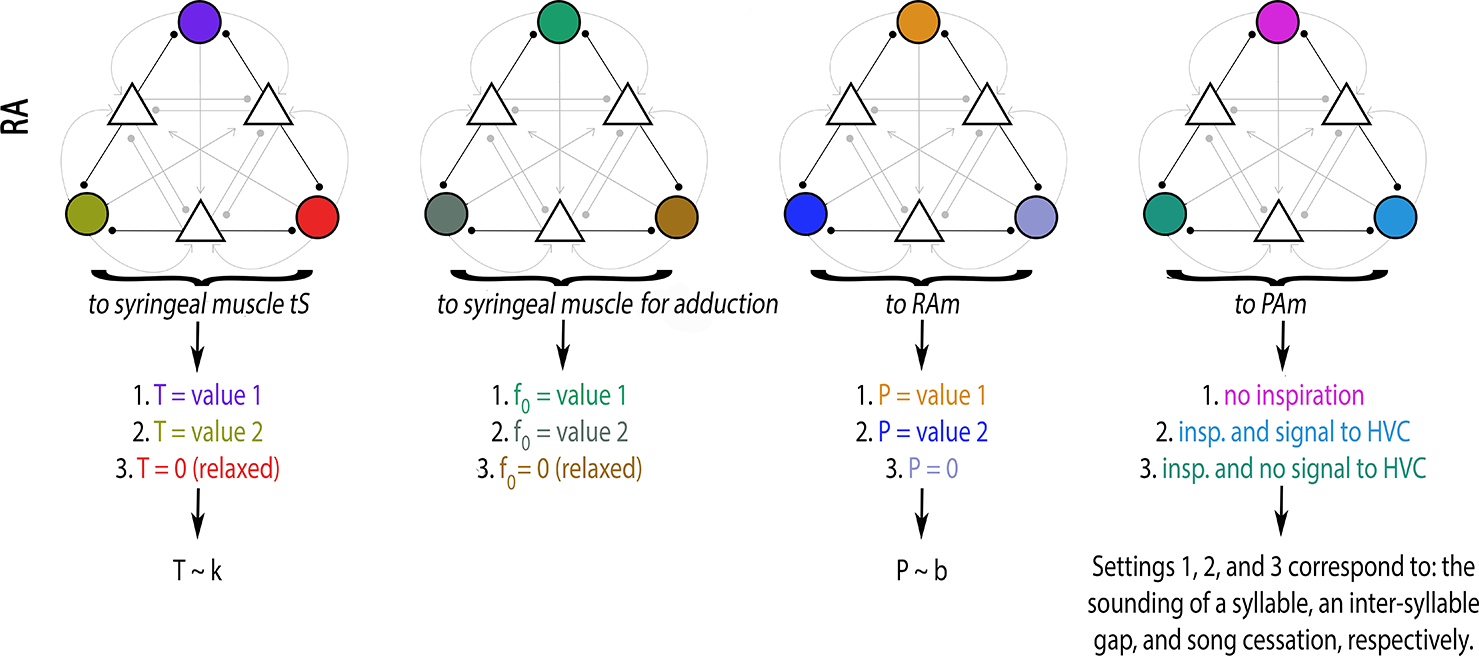} 
\caption{Example assignments for each RA PN in RA, to effect particular values of the driving forces for song.  Within each of the four identical RA structures, each of three RA PNs - in exciting a particular brainstem neuron - codes for a particular order. }
\end{figure}
\begin{multicols}{2}
$f_0$ Value 1 (that is: strongly adduct to prevent phonation\rq\rq; 2) \lq\lq Effect $f_0$ Value 2 (strongly adduct); 3) \lq\lq Set $f_0$ to 0 (that is: to resting, or: permit phonation)\rq\rq.  

For the three RA PNs in the RA structure projecting to RAm (third from left in Figure 6): 1) \lq\lq Effect air sac Pressure Value 1\rq\rq; 2) \lq\lq Effect Pressure Value 2\rq\rq; 3) \lq\lq Set air sac pressure to resting (no expiration)\rq\rq.  

For the three RA PNs in the RA structure projecting to PAm (rightmost in Figure 6): 1) \lq\lq Do not inhale\rq\rq; 2) \lq\lq Inhale and send feedback to HVC (that is: commence the next syllable)\rq\rq; 3) \lq\lq Inhale and do not send feedback (that is: end song)\rq\rq.\footnote{In a larger-scale model, the second command to PAm could be made to contact both the PAM population locked to respiration and the population communicating to Uva, while the third command would recruit only the former population.}

As an example, we take the left-most RA structure in Figure 6 (and 7), which projects to rostral nXIIts.  As described in \textit{Results}, $HVC_{RA}$ PNs 3 and 4 excite RA interneuron 0 so that the only active RA PN is RA 4.  RA 4 directly excites a neuron in rostral nXIIts (Appendix A, Figure 15, bottom panel).  We now say: the activation of that particular neuron in rostral nXIIts incites the order: \lq\lq Effect Tension Value 1\rq\rq\footnote{We are agnostic regarding whether the order is encoded in the synapse between RA PN 4 and the brainstem neuron, or between the brainstem neuron and the ventral syringeal muscle vS.}. 

The complete - and rather arbitrarily-chosen - downstream assignments are shown in Figure 16.  Then Figure 17 shows explicit examples of the type of acoustic structure that may be effected via the first, second, and third $HVC_{RA}$ bursts during song.  That is: the three-paneled Figure 17 is a three-frame movie, where $HVC_{RA}$ PN 3, $HVC_{RA}$ PN 4, and $HVC_{RA}$ PN 5 burst in turn.  Within each frame, black arrows indicate the four specific RA interneurons excited by the $HVC_{RA}$ PN in black at top left.  The bold colorings indicate which RA PN is then permitted to burst during that interval.  Those colorings correspond to the order at bottom whose font is enlarged and bold.  

The frames in Figure 17 go as follows.  The first time bin of Syllable 1 (that is: the firing of $HVC_{RA}$ PN 3) orders: labial tension value $T_ 1$, $f_0$ value 1, air sac pressure value $P_1$, and no inspiration.  The translation to song is: \lq\lq Preventing phonation (adducting the syrinx), effect $T_1$ and $P_2$, and do not inhale.\rq\rq\

The second time bin of Syllable 1 (the firing of $HVC_{RA}$ PN 4) orders: labial tension value $T_1$, $f_0 =$ 0, air sac pressure value $P_2$, and no inspiration.  The translation is: \lq\lq Phonate at tension $T_1$ and pressure $P_2$, and do not inhale.\rq\rq\ 

The third time bin of Syllable 1 (the firing of $HVC_{RA}$ PN 5)  orders: labial tension $T$ $=$ 0, $f_0$ value 2, air sac pressure $P$ $=$ 0, and  inspiration is activated.  The translation is: \lq\lq Stop phonating, inhale, and command the initiation of the subsequent syllable\rq\rq.

These three stages may be compared directly to the stages described in Laje et al. (2002).  Finally, in this four-syllable model the three stages repeat three times, where each set of orders corresponds to the firing of one $HVC_{RA}$ neuron - for a total of 12 song \lq\lq segments\rq\rq.
\end{multicols}
\begin{figure}[H]
  \centering
\includegraphics[width=0.8\textwidth]{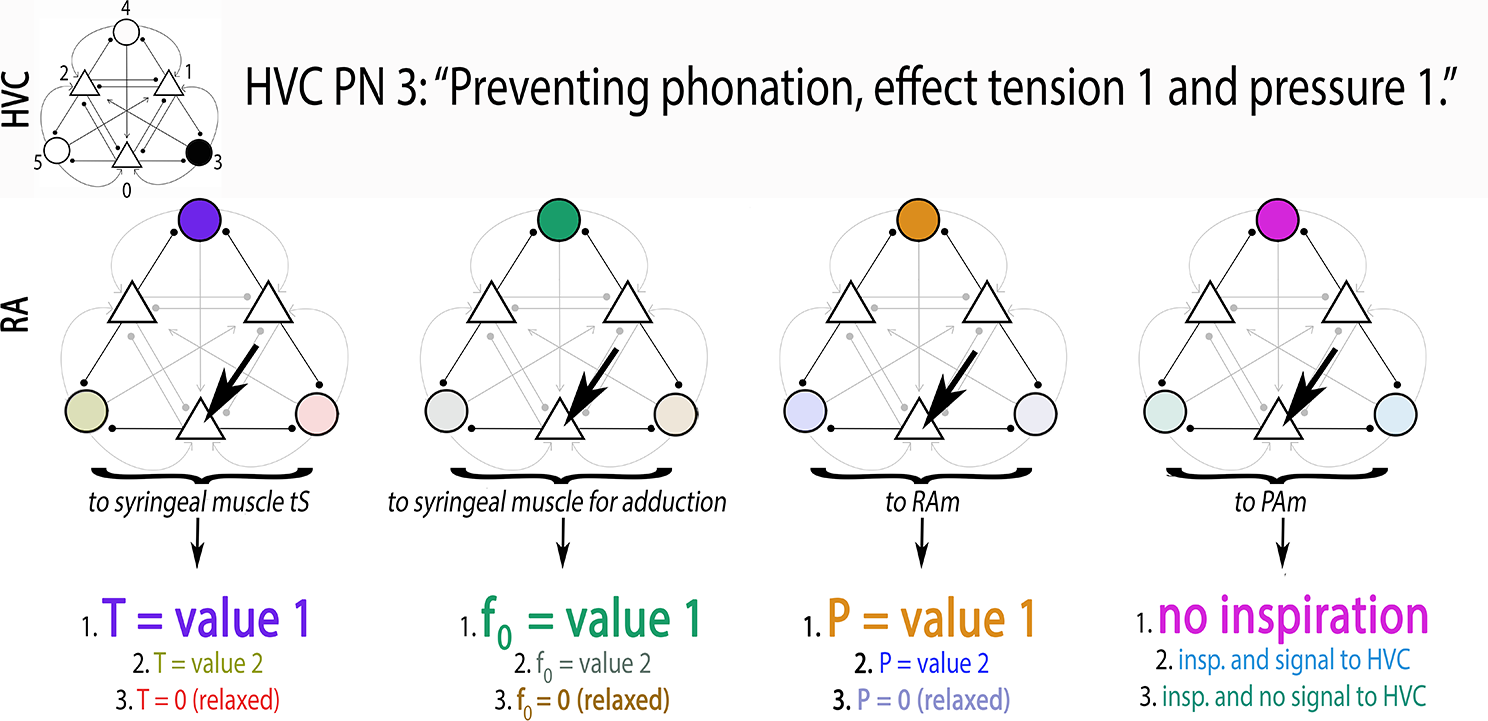}\\
\hrule
\includegraphics[width=0.8\textwidth]{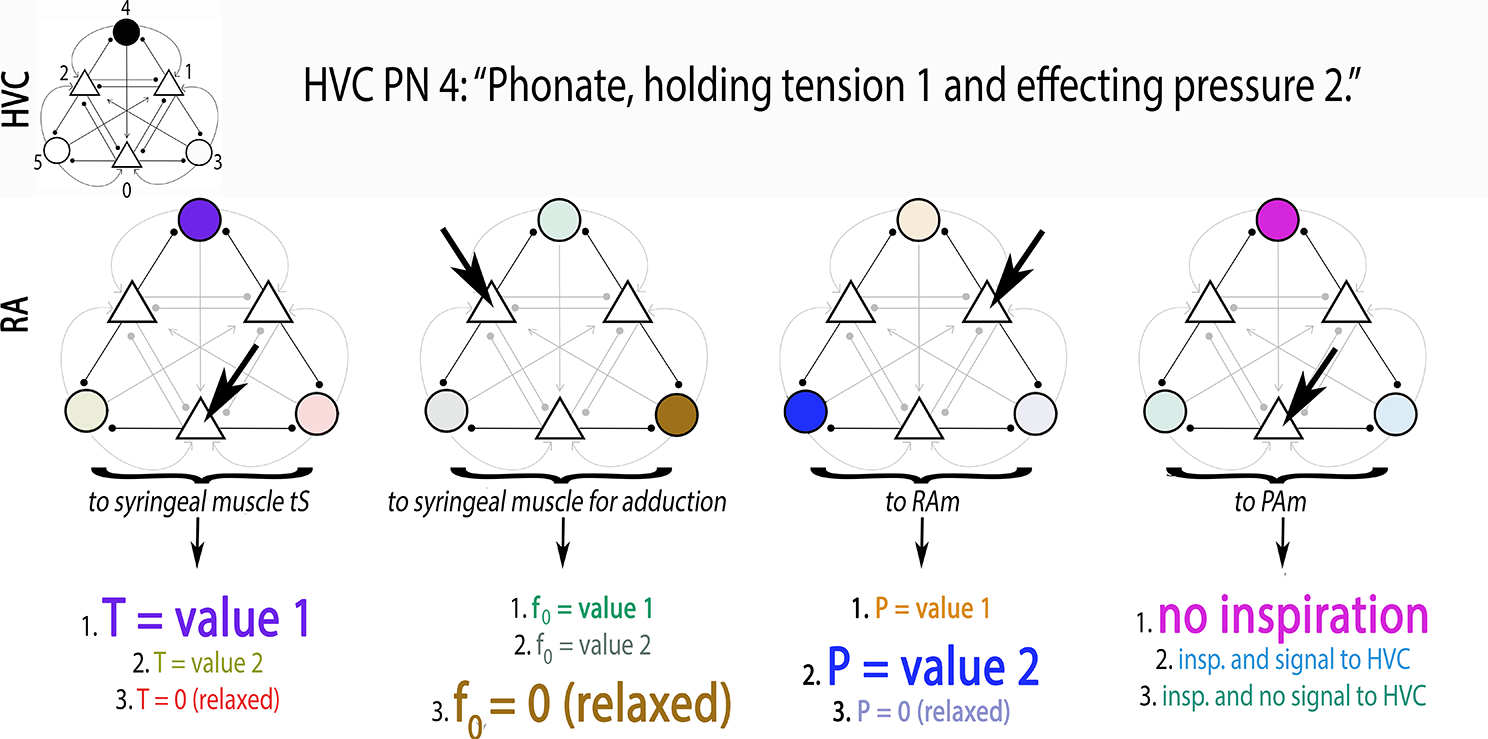}\\
\hrule
\includegraphics[width=0.8\textwidth]{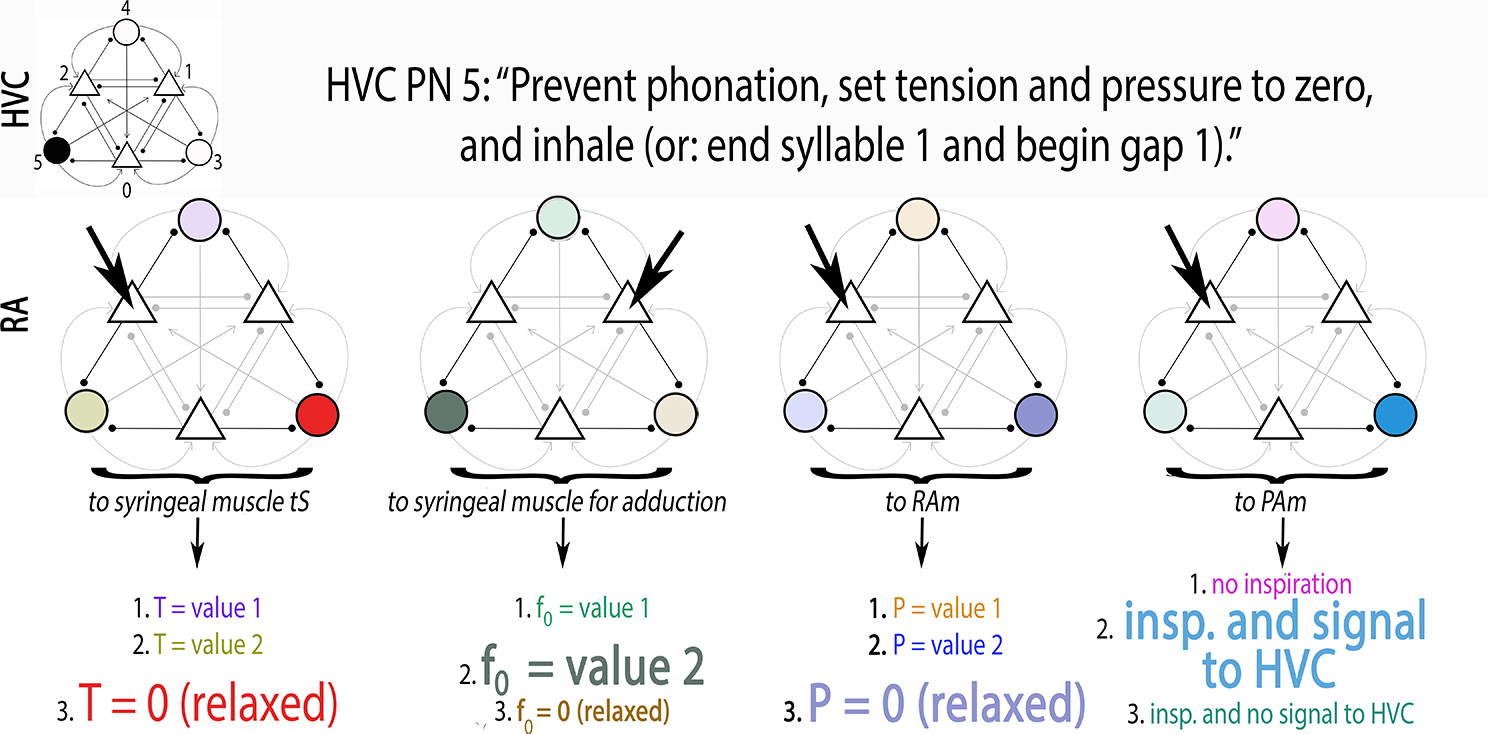}\\
\hrule 
\caption{A three-frame movie illustrating examples of the first three time-bins of a motif - controlled by a burst from $HVC_{RA}$ PN 3, $HVC_{RA}$ PN 4, and $HVC_{RA}$ PN 5, respectively.  In each frame,  black arrows indicate the four particular RA interneurons excited by the $HVC_{RA}$ PN in black at top left.  Then the bold colorings indicate which RA PN is permitted to burst during that interval.  Those colorings correspond to the order at bottom whose font is enlarged and bold.}
\end{figure}
\begin{multicols}{2}
 
\subsection{\textbf{Translating the driving forces into an example spectrogram}}

Extensive modeling has been done of the syrinx and vocal tract, and of the role that the syrinx plays in creating the spectral content of sound (Gardner et al. 2001, Sitt et al. 2008, Amador \& Mindlin 2008, Riede \& Goller 2010, Fee et al. 1998, Elemans et al. 2008, Mindlin 2017a, Mindlin 2017b, Beckers et al. 2003, Jensen et al. 2007).  To create the spectrogram depicted at top in Figure 8, we employed an existing model for the displacement of the labia from resting location (Equation 3.

Perl et al. (2011) then added to this basic model a tract and Helmholtz oscillator to represent the trachea and oropharyngeal cavity, respectively.  This model was used by Amador et al. (2013) to create synthetic sounds that were recognized as BOS playback by HVC of awake birds.  Now, our model driving forces were not derived formally, and for that reason we considered it excessive to filter the air sac pressure through a biophysically-realistic vocal tract in order to produce a spectrogram.  To produce the spectrogram of Figure 8, then, we took the Fourier transform of $x(t)P(t)$, which roughly corresponds to the output acoustic pressure wave (Perl et al. 2011).

\section{Appendix C: Complete equations of motion and parameter values}
\subsection{\textbf{Ion channel gating variables}}

The gating variables $U_i(t) = [n_{f,i}(t), n_{s,i}(t)]$, which govern the fast and slow potassium current, respectively, satisfy:
\begin{align*} 
  \diff{U_i(t)}{t} &= (U_{\infty}(V_i(t)) - U_i(t))/t_{U0}; \\
  U_{\infty}(V_i) &= 0.5 [1 + \tanh((V_i - \theta_{U,i})/\sigma_{U,i})]\\.
\end{align*} 
\noindent
The gating variable for $I_{NaP}$, $m_{inf,i}$, is treated as instantaneous: $0.5 [1 + \tanh((V_i - \theta_{m,i})/\sigma_{m,i})$.  

\subsection{\textbf{Synapse gating variables}}

The synapse gating variables $s_{ij}$, for the synapse entering cell $i$ from cell $j$, evolve as:
\begin{align*}   
  \diff{s_{ij}(t)}{t} &= \alpha(T_{max}(t),V_j(t))[1 - s_{ij}(t)] - \beta s_{ij}(t),
\end{align*}
\noindent
where $\alpha$ is taken to be a function of the maximum neurotransmitter concentration $T_{max}$(t):
\begin{align*}
  \alpha(T_{max}(t),V_j(t)) &= \frac{T_{max}/T_0}{1 + \exp(-(V_{j}(t) - V_{P})/K_{P})}.
\end{align*}   

For the examples in which $T_{max}$ was held static (in order to demonstrate the robustness of the firings between HVC, RA, and brainstem), $T_{max}$ for all connections was held at a value of 2 mMol.  To create the raster plot of Figure 8, the $g_{ij}$-$T_{max}$ relation of Armstrong \& Abarbanel (2016) was invoked.  Within that framework, the  inhibitory-to-inhibitory strengths during active mode are of order 1.0 $\mu$Siemen; see Armstrong \& Abarbane; (2016) for details and for the specific form of $T_{max}(t)$.

\subsection{\textbf{Cellular parameters}}

\textit{HVC}\\
\noindent
The four FSUs are identical and receive a background injected current of 7.88 nA (nano-Amperes).  Table 1 lists the cellular parameters.
\end{multicols}
\begin{table}[H]
\small
\centering
\begin{tabular}{ l|c c c c c c c} \toprule
 \textit{Quantity} & Cell 0 & Cell 1 & Cell 2 & Cell 3 & Cell 4 & Cell 5  & [unit] \\\midrule 
 \textit{$g_{L}$} & 8. & 8. & 8. & 8. & 8. & 8. & [$\mu$S] \\
 \textit{$g_{Na}$} & 20. & 20. & 20. & 20. & 20. & 20. & [$\mu$S] \\
 \textit{$g_{K,s}$} & 9. & 9. & 9. & 9. & 9. & 9. & [$\mu$S]\\
 \textit{$g_{K,f}$} & 4.95. & 4.93 & 4.94 & 4.92 & 4.935 & 4.942 & [$\mu$S] \\ 
 \textit{$E_{L}$} & -80.07 & -80.06 & -80.07 & -80.07 & -80.05 & -80.06 & [mV] \\   
 \textit{$E_{Na}$} & 60. & 60. & 60. & 60. & 60.5 & 60.5 & [mV] \\    
 \textit{$E_{K}$} & -90. & -90. & -90. & -90. & -90.5 & -90.5 & [mV] \\    
 \textit{$\theta_{m}$} & -20. & -20. & -20. & -20. & -20. & -20. & [mV]\\
 \textit{$\sigma_{m}$} & 30. & 30. & 30. & 30. & 30. & 30. & [mV]\\
 \textit{$\theta_{n,f}$} & -25. & -25. & -25. & -25. & -25. & -25. & [mV]\\
 \textit{$\sigma_{n,f}$} & 10. & 10. & 10. & 10. & 10. & 10. & [mV]\\
 \textit{$t_{n,f}$} & 0.152 & 0.152 & 0.152 & 0.152 & 0.152 & 0.152 & [ms]\\

 \textit{$\theta_{n,s}$} & -20. & -20. & -20. & -20. & -20. & -20. & [mV]\\
 \textit{$\sigma_{n,s}$} & 5. & 5. & 5. & 5. & 5. & 5. & [mV]\\
 \textit{$t_{n,s}$} & -20. & -20. & -20. & -20. & -20. & -20. & [ms]\\
 \textit{C} & 1. & 1. & 1. & 1. & 1. & 1. & [$\mu$F] \\\bottomrule
\end{tabular}\\
\caption{Parameter values for six cells in an FHU of HVC.  All four FHUs in the model presented in this paper are identical in terms of these values.  Units: mV, millivolts; ms, milliseconds; $\mu$F, micro-Farads; $\mu$S, micro-Siemens.} 
\end{table}
\begin{multicols}{2}

\noindent
\textit{RA}\\
\noindent
The four six-neuron structures in RA are identical, and each receives a background current of 7.81 nA (99 per cent the value of background current injected to HVC).  The cellular parameters for the RA structures are identical to those listed above for an FHU in HVC, except that the leak reversal potentials $E_L$ for the interneurons (Cells 0, 1, and 2) are: -85.0 mV, rather than -80.0 mV for the excitatory PNs.  \\

\noindent
\textit{Brainstem}\\
\noindent
The cellular parameters for each of the four brainstem neurons are identical to those of RA PN 5 (and hence also $HVC_{RA}$ PN 5).
\begin{table}[H]
\small
\centering
\begin{tabular}{ l|c c c c c c c} \toprule
 \textit{Cell} & 0 & 1 & 2 & 3 & 4 & 5  \\\midrule 
 0 & 0. & -83. & -83.3 & 0. & 0. & 0.  \\
 1 & -82.7 & 0. & -82.5 & 0. & 0. & 0.  \\
 2 & -83.2 & -82.9 & 0. & 0. & 0. & 0. \\
 3 & -83.0 & -83.0 & -83.0 & 0. & 0. & 0. \\
 4 & -83.0 & -83.0 & -83.0 & 0. & 0. & 0. \\
 5 & -83.0 & -83.0 & -83.0 & 0. & 0. & 0.\\\bottomrule 
\end{tabular}\\
\caption{Synaptic reversal potentials $E_{ij}$, for the synapse entering cell $i$ from cell $j$.  Units: mV.} 
\end{table}

\subsection{\textbf{Synapse parameters}}

The parameters $E_{ij}$, $\alpha_{ij}$, and $\beta_{ij}$, for each six-cell structure in both HVC and RA, are listed in Tables 2, 3, and 4, respectively.
\begin{table}[H]
\small
\centering
\begin{tabular}{ l|c c c c c c c} \toprule
 \textit{Cell} & 0 & 1 & 2 & 3 & 4 & 5  \\\midrule 
 0 & 1. & 1. & 1.1 & 2.2 & 2.2 & 2.2  \\
 1 & 1.05 & 1. & 1.9 & 2.2 & 2.2 & 2.2  \\
 2 & 1.2 & 1.8 & 1. & 2.2 & 2.2 & 2.2 \\
 3 & 1.0 & 1.0 & 1.0 & 2. & 2. & 2. \\
 4 & 1.0 & 1.0 & 1.0 & 2. & 2. & 2. \\
 5 & 1.0 & 1.0 & 1.0 & 2. & 2. & 2.\\\bottomrule 
\end{tabular}
\caption{Parameters $\alpha_{ij}$, for the synapse entering cell $i$ from cell $j$.  Units: $mMol^{-1} ms^{-1}$.} 
\end{table}
\begin{table}[H]
\small
\centering
\begin{tabular}{ l|c c c c c c c} \toprule
 \textit{Cell} & 0 & 1 & 2 & 3 & 4 & 5  \\\midrule 
 0 & 0. & 0.18 & 0.181 & 0.38 & 0.38 & 0.38  \\
 1 & 0.182 & 0. & 0.179 & 0.38 & 0.38 & 0.38  \\
 2 & 0.178 & 0.183 & 0. & 0.38 & 0.38 & 0.38 \\
 3 & 0.18 & 0.18 & 0.18 & 0.38 & 0.38 & 0.38 \\
 4 & 0.18 & 0.18 & 0.18 & 0.38 & 0.38 & 0.38 \\
 5 & 0.18 & 0.18 & 0.18 & 0.38 & 0.38 & 0.38 \\\bottomrule 
\end{tabular}
\caption{Parameters $\beta_{ij}$, for the synapse entering cell $i$ from cell $j$.  Units: $ms^{-1}$.} 
\end{table}

For the connections between RA and brainstem neurons, the values of $E_{ij}$, $\alpha_{ij}$, and $\beta_{ij}$ were taken for an excitatory-to-excitatory connection.  We arbitrarily chose the values for a synapse entering cell 3 from cell 5.\\

\noindent
\textit{For an HVC FHU during active mode}\\

\noindent
For the $g_{ij}$-$T_{max}$ relation governing the inhibitory-to-inhibitory connections (among Cells 0, 1, and 2), the parameters $V_p$ and $K_p$ are:
\setlength{\tabcolsep}{1pt}
\begin{table}[H]
\small
\centering
\begin{tabular}{ l c c c c} \toprule
 & Cell 0 & 1 & 2 & [unit]\\\midrule 
 $V_p$ & 2. & 2.01 & 2.03 & [mV]  \\
 $K_p$ & 5.0 & 5.01 & 4.8 & [mV]  \\\bottomrule
\end{tabular}
\end{table}
\noindent
For all other connections: $V_p = 2.$ mV; $K_p = 5.$ mV; $T_{max} = $ 2.0 mM.\\

\noindent
\textit{Strengths of static synaptic connections}\\

\noindent
In HVC: the excitatory-to-inhibitory, and inhibitory-to-excitatory synapse strengths are: 1.0 and 2.5 $\mu$S, respectively.  In RA: within each of the four RA structures (of Figure 5), the inhibitory-to-excitatory connections - which are critical for the model operation - are 10.0 $\mu$S.  For cross-connectivity from HVC-to-RA and from RA-to-brainstem: the synapse strengths are 10 $\mu$S.
\end{multicols}

\end{document}